\documentclass[journal]{IEEEtran}
\usepackage{mathrsfs}
\usepackage{pifont}
\usepackage{bbding}
\usepackage{tipa}
\usepackage{algorithm}
\usepackage{algorithmic}
\usepackage{amsmath,epsfig}
\usepackage{stmaryrd}
\usepackage{amssymb}
\usepackage{amsfonts}
\usepackage{epic}
\usepackage{graphicx}
\ifCLASSINFOpdf
\else
\fi
\begin{document}\newcounter{fofo}
\newtheorem{theorem}{Theorem}
\newtheorem{proposition}{Proposition}
\newtheorem{definition}{Definition}
\newtheorem{lemma}{Lemma}
\newtheorem{corollary}{Corollary}
\newtheorem{remark}{Remark}
\newtheorem{construction}{Construction}

\newcommand{\supp}{\mathop{\rm supp}}
\newcommand{\sinc}{\mathop{\rm sinc}}
\newcommand{\spann}{\mathop{\rm span}}
\newcommand{\essinf}{\mathop{\rm ess\,inf}}
\newcommand{\esssup}{\mathop{\rm ess\,sup}}
\newcommand{\Lip}{\rm Lip}
\newcommand{\sign}{\mathop{\rm sign}}
\newcommand{\osc}{\mathop{\rm osc}}
\newcommand{\R}{{\mathbb{R}}}
\newcommand{\Z}{{\mathbb{Z}}}
\newcommand{\C}{{\mathbb{C}}}
%
% paper title
% can use linebreaks \\ within to get better formatting as desired

%=======================================================title==============================================================================
%\title{Adaptive Subcarrier Pairing and Power Allocation with Limited Feedback for the OFDM Relay Networks}
\title{Limited Feedback based Adaptive Power Allocation and Subcarrier Pairing for OFDM DF Relay Networks with Diversity}
% author names and affiliations
% use a multiple column layout for up to three different
% affiliations
%=======================================================author information=================================================================
%\author{\IEEEauthorblockN{Yong~Liu\IEEEauthorrefmark{1},
%Wen~Chen\IEEEauthorrefmark{1},
%Xiaopeng~Huang\IEEEauthorrefmark{2},
%%Montgomery Scott\IEEEauthorrefmark{3} and
%%Eldon Tyrell\IEEEauthorrefmark{4}}
%\IEEEauthorblockA{\IEEEauthorrefmark{1}Department of Electronic Engineering\\
%Shanghai Jiaotong University, Shanghai, 200240\\ Email: \{yongliu1982, wenchen\}@sjtu.edu.cn}.
%\IEEEauthorblockA{\IEEEauthorrefmark{2}Electrical and Computer engineering department\\Stevens Institute of Technology, New Jersey, USA\\
%Email: xhuang3@stevens.edu}}}
%\IEEEauthorblockA{\IEEEauthorrefmark{3}Starfleet Academy, San Francisco, California 96678-2391\\
%Telephone: (800) 555--1212, Fax: (888) 555--1212}
%\IEEEauthorblockA{\IEEEauthorrefmark{4}Tyrell Inc., 123 Replicant Street, Los Angeles, California 90210--4321}}

\author{{Yong~Liu,}
        and~Wen~Chen,~\IEEEmembership{Senior Member,~IEEE}
        % <-this % stops a space
\thanks{Copyright (c) 2012 IEEE. Personal use of this material is
permitted. However, permission to use this material for any other
purposes must be obtained from the IEEE by sending a request to
pubs-permissions@ieee.org.}
\thanks{Manuscript received November 29, 2011; revised March 6, 2012; accepted
April 30, 2012. The review of this paper was coordinated by Prof.
Michel Yacoub.}
\thanks{The authors are with the Department of Electronic Engineering,
Shanghai Jiao Tong University, Shanghai, China, and SKL for
ISN, Xidian University, China. e-mail: \{yongliu1982, wenchen\}@sjtu.edu.cn}% <-this % stops a space
\thanks{This work is supported by national 973 project
\#2012CB316106, by NSFC \#60972031 and \#61161130529, by national
973 project \#2009CB824900, and by national key laboratory project
\#ISN11-01.}}

% conference papers do not typically use \thanks and this command
% is locked out in conference mode. If really needed, such as for
% the acknowledgment of grants, issue a \IEEEoverridecommandlockouts
% after \documentclass

% for over three affiliations, or if they all won't fit within the width
% of the page, use this alternative format:
%
%\author{\IEEEauthorblockN{Michael Shell\IEEEauthorrefmark{1},
%Homer Simpson\IEEEauthorrefmark{2},
%James Kirk\IEEEauthorrefmark{3},
%Montgomery Scott\IEEEauthorrefmark{3} and
%Eldon Tyrell\IEEEauthorrefmark{4}}
%\IEEEauthorblockA{\IEEEauthorrefmark{1}School of Electrical and Computer Engineering\\
%Georgia Institute of Technology,
%Atlanta, Georgia 30332--0250\\ Email: see http://www.michaelshell.org/contact.html}
%\IEEEauthorblockA{\IEEEauthorrefmark{2}Twentieth Century Fox, Springfield, USA\\
%Email: homer@thesimpsons.com}
%\IEEEauthorblockA{\IEEEauthorrefmark{3}Starfleet Academy, San Francisco, California 96678-2391\\
%Telephone: (800) 555--1212, Fax: (888) 555--1212}
%\IEEEauthorblockA{\IEEEauthorrefmark{4}Tyrell Inc., 123 Replicant Street, Los Angeles, California 90210--4321}}

% use for special paper notices
%\IEEEspecialpapernotice{(Invited Paper)}

% make the title area
\maketitle

%=======================================================abstract=================================================================
\begin{abstract}
%\boldmath
A limited feedback based dynamic resource
allocation algorithm is proposed for a relay cooperative network with Orthogonal Frequency
Division Multiplexing (OFDM) modulation. A communication model where one source node
communicates with one destination node assisted by one half-duplex
Decode-and-Foward (DF) relay is considered in this paper. We first consider the \emph{selective} DF
scheme, in which some relay subcarriers will keep idle if they are not advantageous to forward the
received symbols. Furthermore, we consider the
\emph{enhanced} DF scheme where the idle subcarriers are used to
transmit new messages at the source. We aim to maximize the system instantaneous
rate by jointly optimizing power allocation and subcarrier pairing
on each subcarrier based on the Lloyd algorithm. Both sum
and individual power constraints are considered. The joint
optimization turns out to be a mixed integer programming problem.
We then transform it into a convex optimization by continuous
relaxation, and achieve the solution in the dual domain.
%We present an
%adaptation of Lloyd¡¯s algorithm to construct a codebook to quantize
%the optimal power allocation and SP vectors subject to the amount of
%feedback.
The performance of the proposed joint resource allocation algorithm is verified by simulations. We find that the proposed scheme outperforms the existing methods in various channel
conditions. We also observe
that only a few feedback bits can achieve most of the
performance gain of the perfect CSI based resource allocation
algorithm at different levels of SNR.
%
%A negligible performance loss can be achieved with just
%a few feedback bits
\end{abstract}


% IEEEtran.cls defaults to using math in the Abstract.
% This preserves the distinction between vectors and scalars. However,
% if the conference you are submitting to favors bold math in the abstract,
% then you can use LaTeX's standard command \boldmath at the very start
% of the abstract to achieve this. Many IEEE journals/conferences frown on
% math in the abstract anyway.

% no keywords

% For peer review papers, you can put extra information on the cover
% page as needed:
% \ifCLASSOPTIONpeerreview
% \begin{center} \bfseries EDICS Category: 3-BBND \end{center}
% \fi
%
% For peerreview papers, this IEEEtran command inserts a page break and
% creates the second title. It will be ignored for other modes.
\IEEEpeerreviewmaketitle

%=======================================================section1 introduction=======================================================
\section{Introduction}\label{sec:1}
% no \IEEEPARstart
Considering the limited budget of transmit power and hardware complexity,
cooperative relaying has recently attracted a lot of research interests, which is employed
to exploit spatial diversity, combat
wireless channel fading and extend coverage without antenna arrays
\cite{IEEEconf:1}-\cite{IEEEconf:2}. For example, IEEE 802.16
currently integrates relays for multihop communications
\cite{IEEEconf:4}. Two main relay strategies have been adopted in
such scenarios: Amplify-and-Forward (AF) and Decode-and-Forward
(DF).
The AF relay amplifies and retransmits the received signal without decoding, while
the latter re-encodes the received signal before retransmission.

%Orthogonal Frequency Division Multiplexing (OFDM) is a promising
%technique to mitigate the frequency selectivity and inter-symbol
%interference.
%
%In AF scheme, the received signal is amplified and then
%retransmitted by the relay without performing decoding, while
%the latter decodes the received signal and re-encodes it for
%retransmission.

Orthogonal Frequency Division Multiplexing (OFDM) is a technique to mitigate the frequency selectivity and inter-symbol
interference with its inherent robustness against frequency-selective fading~\cite{IEEEconf:22}.
%By
%employing IFFT and FFT processing, a frequency-selective fading
%channel can be converted into parallel flat-fading ones.
Because of its potential for high spectral efficiency, OFDM-based
relaying
%OFDM has become a
%promising candidate for the physical layer in $4G$ cellular
%system~\cite{IEEEconf:22}.
offers a more promising perspective in improving
system performance.

Power allocation is always critical in wireless networks
due to the limited budget of transmit power. It has been widely discussed in the context of both single-carrier and multi-carrier relaying
channels~\cite{IEEEconf:23}-\cite{IEEEconf:11}. We have proposed limited feedback based power allocation algorithms for a single-carrier relaying channel and a multi-carrier based relaying model in~\cite{IEEEconf:23} and~\cite{IEEEconf:28} respectively. In \cite{IEEEconf:6}, Ahmed \emph{et al.} propose a power
control algorithm for AF relaying with limited feedback. Then they study the rate and power
control to improve the throughput gain of DF in~\cite{IEEEconf:5}.
%  is proposed by
%
% propose  channel in~\cite{IEEEconf:5}, and
% allocation to minimize the outage probability with AF
%relay in .
%Similarly, optimal power allocation that
%maximizes system capacity of relay channel is proposed in
%\cite{IEEEconf:7}.
On the other hand, power allocation for OFDM-based
relaying is also extensively studied. Authors in~\cite{IEEEconf:8} investigate the
power allocation for an OFDM based AF relaying by separately
optimizing the source and relay powers. In \cite{IEEEconf:9}, the same
authors propose a power allocation scheme for MIMO-OFDM relay system in the same way.
%Unlike AF relaying in
%\cite{IEEEconf:9},
In~\cite{IEEEconf:10}, Ying \emph{et al.} work on the similar problem but for DF
relaying OFDM systems. Ma \emph{et al.} introduce power loading algorithms to minimize the
transmit power for OFDM based AF and \emph{selective} DF modes with respect to various power-constraint conditions in~\cite{IEEEconf:11}.

Due to the independent fading on each subcarrier in each hop, subcarrier
pairing is employed in OFDM power allocation to further improve system
performance~\cite{IEEEconf:25}-\cite{IEEEconf:42}. Most works in
literature focus only on relay models without diversity.
%In~\cite{IEEEconf:24} and~\cite{IEEEconf:25}, optimal power
%allocation for OFDM with DF relaying and fixed subcarrier pairing
%was proposed.
A sorted subcarrier pairing scheme is proposed in~\cite{IEEEconf:25}. The authors determine the pairing sequence by ordering the the source-relay (SR) subcarriers and the
relay-destination (RD) subcarriers, respectively, according to the channel
gains. Authors in~\cite{IEEEconf:26} prove that the sorted pairing method is optimal for
both DF and AF relaying without the source-destination (SD) link.
%However, how to optimally pairing subcarriers in relay models with
%diversity is still unsolved.
Authors in~\cite{IEEEconf:40} jointly optimize channel
pairing, channel-user assignment, and power allocation in a
multiple-access system by a polynomial-time algorithm
based on continuous relaxation and dual minimization.
Wang \emph{et al.} in~\cite{IEEEconf:41} propose a joint subcarrier pairing and power allocation algorithm for an OFDM two-hop relay system with separate power constraints, and find the solution by separately considering the subcarrier pairing and the power allocation. Authors in~\cite{IEEEconf:42} investigate optimal subcarrier assignment and power allocation schemes for multi-user multi-relay model, and obtain the optimal subcarrier and power allocation policy in a quasi-closed
form.

%introduce a pairing method for DF with
%diversity, but subcarrier pairing and power allocation are
%independently performed, and hard to be realized via limited
%feedback.

%Authors in
%\cite{IEEEconf:8} propose SCP for OFDM AF when the SD link and
%destination combining are present.

Resource allocation utilizing channel state information (CSI) can yield
significant performance improvement~\cite{IEEEconf:23}-\cite{IEEEconf:28},\cite{IEEEconf:12}-\cite{IEEEconf:14}.
Tremendous innovation that realize instantaneous channel adaptation
is to use feedback whose history may trace back to
Shannon \cite{IEEEconf:15}. It is proved that with perfect CSI at source, the error and
capacity performance are significantly
better than that without
CSI~\cite{IEEEconf:14,IEEEconf:16}. Some research have been carried out to achieve the performance gain based on
limited feedback, since perfect CSI at source is always impractical.
%for too much
%%obtaining the full knowledge of CSI
%%at the transmitters. Obviously this needs
%extra spectral
%resource.
One can either send back a quantized CSI or
%send the
quantized power allocation vectors~\cite{IEEEconf:6}, or
%sends back
the index of the best vector in a power allocation codebook shared
by all nodes~\cite{IEEEconf:18}-\cite{IEEEconf:19}. These works are mostly
studied in point-to-point MIMO and OFDM systems. Only
a few works exist on OFDM relay networks. Authors
in~\cite{IEEEconf:29} investigate the power allocation issue
for a single OFDM AF relay network with limited feedback. They construct the codebook based on the Lloyd algorithm.
%present a based
%codebook construction to  subject to the amount of feedback. In contrast,
Similarly, Zhang \emph{et al.} introduce the same idea into DF model in~\cite{IEEEconf:30}.
%, where
%they conclude that the performance of the limited feedback scheme
%lies between the perfect CSI scheme and the scheme without CSI.

%Much work has also been reported in the literature on subcarrier
%pairing for the OFDM based relay
%networks~\cite{IEEEconf:24}-\cite{IEEEconf:27},
%In~\cite{IEEEconf:24} and~\cite{IEEEconf:25}, optimal power
%allocation for OFDM with DF relaying and fixed source and relay
%subcarrier pairing was proposed. Both the total and individual power
%constraints are considered in these works. In~\cite{IEEEconf:26}
%and~\cite{IEEEconf:27}, both power allocation and subcarrier pairing
%were considered for OFDM systems with relaying under the total power
%constraint, in which the power allocation and subcarrier pairing
%were optimized separately. %OSP was proposed in~\cite{IEEEconf:28}
%%and it focused on only the total power constraint.
%To the best of our knowledge, using limited feedback in such a joint
%optimization inclusive of the destination combining has not been
%discussed in the literature.

In view of the lack of joint optimization of power allocation and
subcarrier pairing for OFDM relay systems with diversity based on limited feedback, we aim to solve this
problem in this paper.
% and destination combining of signals from
%the source and the relay.
This work is developed based on our previous works~\cite{IEEEconf:23} and~\cite{IEEEconf:28}.
We present a limited feedback based joint power allocation and
subcarrier pairing for a \emph{selective} OFDM DF relay network
under different levels of quantized CSI feedback. In our feedback
scheme, we construct a codebook based on an iterative Lloyd algorithm
with a modified distortion measure.
% with a joint power constraint, and employ the Lloyd
%algorithm to construct the optimal codebook.
The joint optimization problem is formulated as a mixed integer
programming problem which is hard to solve. We transform it into a
convex problem by continuous relaxation~\cite{IEEEconf:35},
\cite{IEEEconf:43}, and solve it in the dual domain instead. In our
simulation, we observe that the duality gap virtually turns out to
be zero when the number of subcarriers is reasonably large, which is
consistent with that observed in~\cite{IEEEconf:43}
and~\cite{IEEEconf:44}.

%Taking the same assumption as
%in~\cite{IEEEconf:10},~\cite{IEEEconf:31},  For
%example,

We then relax the constraint that only the relay can transmit in the
relaying phase. When the relay does not transmit on some
subcarriers, we employ the \emph{enhanced} DF which allows the source
to transmit new messages on these idle subcarriers. Then we extend
the joint optimization problem for \emph{selective} DF to that for
\emph{enhanced} DF under both sum power constraint and individual
power constraints. It is shown that the extra direct-link
transmission leads to a remarkable rate enhancement in the
simulation. Besides, some existing schemes such as the conventional
uniform power allocation without subcarrier pairing (UPA w/o SP),
the optimal power allocation without subcarrier pairing (OPA w/o
SP), and the uniform power allocation with subcarrier pairing (UPA
with SP) are compared with the proposed algorithm. Simulation
results demonstrate that the proposed algorithm outperforms the
existing ones. We also find that a negligible performance loss can
be achieved with just a few feedback bits at different levels of
SNR.

The remainder of the paper is organized as follows. The system model is introduced in Section~\setcounter{fofo}{2}\Roman{fofo}. In Section~\setcounter{fofo}{3}\Roman{fofo}, we solve the
joint optimization problem for \emph{selective} DF relay networks, and propose a limited feedback based resource allocation algorithm. In Section~\setcounter{fofo}{4}\Roman{fofo}, we solve the optimization
problem for \emph{enhanced} DF relay networks, and then consider the joint optimization under individual power constraints. Simulations are performed
in Section~\setcounter{fofo}{5}\Roman{fofo} to verify the performance of the proposed algorithm.
Finally the conclusions are drawn in Section~VI.

% You must have at least 2 lines in the paragraph with the drop letter
% (should never be an issue)
%=======================================================section2=====================================================================
\section{System Model}\label{sec:2}

The scenario of three nodes DF diversity model is
considered, where one source communicates with one
destination assisted by one half-duplex relay as shown in Fig.~1.
%
%The
%relay does not produce its own data and only assists in the
%communication between the source and destination nodes.
The channel on each hop is divided into \emph{N} subcarriers.
Communication takes place in two phases. The source broadcasts its signal in
the listening phase, while the relay and the destination listen. The relay decodes and
forwards in the relaying phase.
%a total \emph{W} bandwidth is divided into \emph{N} independent
%sub-channels. Communication between the source and destination takes
%place in two phases, i.e., the listening phase and the relaying
%phase. In the listening phase, the source broadcasts the
%information, and the relay and the destination listen; while in the
%relaying phase, the relay transmits a re-encoded version of the
%received signal.
It is assumed that each subcarrier in the listening
phase is paired with one subcarrier in the
relaying phase. So the number of subcarrier pairs is $N$. We utilize
$\mbox{SP}(m,n)$ to denote the subcarrier $m$ in the
listening phase pairing with the subcarrier $n$ in the relaying
phase. For subcarrier pair $\mbox{SP}(m,n)$, it might not be the
actual pair participating in communication. If $\mbox{SP}(m,n)$
actually participates in communication, it is said to be "selected".
%
%Finally the destination tries to decode
%message based on all the signals received during the two phases.
\begin{figure}[!t]
\centering
\includegraphics[width=3.2in,angle=0]{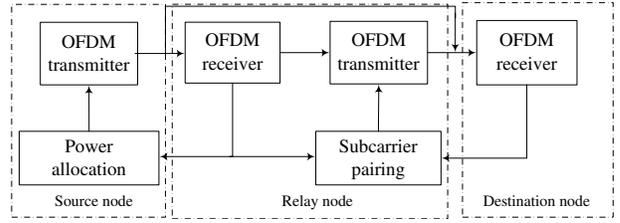}
\caption{System block diagram of OFDM cooperative diversity model
with limited feedback} \label{fig1}
\end{figure}
%When the channel model is associated with $\mbox{SP}(m,n)$, we a
We denote $h^{m}_{SD}$, $h^{m}_{SR}$ and $h^{n}_{RD}$ as channel coefficients of the \emph{m}th subcarrier of source-destination and source-relay, and the \emph{n}th subcarrier
of relay-destination respectively.
%
%
%Assume that
%channels experienced by the \emph{m}th subcarrier of
%source-destination and source-relay, and the \emph{n}th subcarrier
%of relay-destination are attenuated by fading coefficient
%$h^{m}_{SD}$, $h^{m}_{SR}$ and $h^{n}_{RD}$ respectively.
%With the
%channel model associated with
For a potential $\mbox{SP}(m,n)$, the source transmits symbol $s_{m}$ over subcarrier
$m$ with power $P^{m,n}_{S}$ in the listening phase, the received signals at the relay and destination are respectively given by
\begin{equation}
\begin{split}
&y_{rm}=\sqrt{P^{m,n}_{S}}h^{m}_{SR}s_{m}+z_{rm},\\
    &y^{(1)}_{dm}=\sqrt{P^{m,n}_{S}}h^{m}_{SD}s_{m}+z^{(1)}_{dm},
\end{split}
\end{equation}
where
%$y^{(1)}_{dm}$ and $y_{rm}$ are respectively the received
%signals at the destination and relay nodes, and
$z^{(1)}_{dm}\thicksim\mathcal{CN}(0,\sigma_{d}^{2})$ and
$z_{rm}\thicksim\mathcal{CN}(0,\sigma_{r}^{2})$ are the additive
noises at the relay and the destination, respectively.
%, and
%$P^{m,n}_{S}$ is the source power on the $m$th subcarrier in the
%listening phase associated with $\mbox{SP}(m,n)$.
In the relaying
phase, the relay transmits the re-encoded signal $\hat{s}_{m}$ with power $P^{m,n}_{R}$ on
$n$th subcarrier, the received signal at the destination is
\begin{equation}
    y^{(2)}_{dn}=\sqrt{P^{m,n}_{R}}h^{n}_{RD}\hat{s}_{m}+z^{(2)}_{dn},
\end{equation}
where $z^{(2)}_{dn}\thicksim\mathcal{CN}(\mu,\sigma_{d}^{2})$ is the
additive noise at the destination in the relaying phase.
%and
%$P^{m,n}_{R}$ is the relay power on  in the relaying
%phase associated with $\mbox{SP}(m,n)$.
%$P^{m,n}_{S}$ and
%$P^{m,n}_{R}$ have to be carefully designed to optimize some
%performance criteria such as capacity.

Let
%$R^{m,n}$ be the achievable rate concerning SP(m,n), we use
$\lambda^{m}_{SR}=\frac{|h^{m}_{SR}|^{2}}{\sigma_{r}^{2}}$,
$\lambda^{n}_{RD}=\frac{|h^{n}_{RD}|^{2}}{\sigma_{d}^{2}}$, and
$\lambda^{m}_{SD}=\frac{|h^{m}_{SD}|^{2}}{\sigma_{d}^{2}}$ denote
the normalized channel gains respectively. Depending on whether the
relay is helpful, each subcarrier pair $\mbox{SP}(m,n)$ may work in
either the relaying mode or the idle mode in a \emph{selective} DF
relay~\cite{IEEEconf:31}. For an $\mbox{SP}(m,n)$, the relay
forwards the message $\hat{s}_{m}$ on subcarrier $n$ in the relaying
phase when it works in the relaying mode;
%in which we denote $(m,n)\in S_{r}$,
while in the idle mode, the relay does not forward
$(P^{m,n}_{R}=0)$, and the $s_{m}$ is transmitted to destination by
the SD link in the listening phase only.
%in which we denote $(m,n)\in S_{s}$.
Then the end-to-end rate achieved by $\mbox{SP}(m,n)$ during the two
phases is given by
%\begin{displaymath}
%\mathbf D=\begin{cases}
%\frac{1}{1+\text{SNR}\lambda_i} & i\leq\text{rank}(\mathbf {\tilde H})\\
%1 & i>\text{rank}(\mathbf {\tilde H})
%\end{cases}
%\end{displaymath}
\begin{displaymath} \tag{3}
R^{m,n}=\begin{cases}
\;\frac{1}{2}\log_{2}\left(1+P^{m,n}_{S}\lambda^{m}_{SD}\right), \;\;\;\quad\quad\, \mbox{idle\, mode},\\
\;\frac{1}{2}\min\{\log_{2}\left(1+P^{m,n}_{S}\lambda^{m}_{SD}+P^{m,n}_{R}\lambda^{n}_{RD}\right),\\
\qquad\quad\log_{2}\left(1+P^{m,n}_{S}\lambda^{m}_{SR}\right)\},\;\;\;
\mbox{relaying \,mode}.
\end{cases}
\end{displaymath}
Authors in~\cite{IEEEconf:10} and~\cite{IEEEconf:31} present a criterion to decide the
working mode of $\mbox{SP}(m,n)$, that is, using relay is
advantageous when \setcounter{equation}{3}
\begin{equation}
\min\left\{\lambda^{m}_{SR},\lambda^{n}_{RD}\right\}>\lambda^{m}_{SD},
\end{equation}
in \emph{selective} DF mode. Otherwise the relay keeps idle on the
subcarrier $n$ in the relaying phase for $\hat s_{m}$.

%=======================================================section3=====================================================================

\section{Limited Feedback Based Optimal Resource Allocation}\label{sec:3}
In this section, we analyze the joint optimization of power allocation
and subcarrier pairing for \emph{selective} DF based on the limited feedback.
The optimization problem is formulated first,
% try to relax
%the constraint
and then solved in the dual domain.
%Then we discuss the
%algorithm design issues.

\subsection{Optimization Problem Formulation}
%Without loss of generality,
Let $P^{m,n}=P^{m,n}_{S}+P_{R}^{m,n}$ for the SP$(m,n)$. We first
consider the rate $R^{m,n}$ in the relaying mode. Then the sum rate
is maximized when
\begin{equation}
\log_{2}\left(1+P^{m,n}_{S}\lambda^{m}_{SR}\right)=\log_{2}\left(1+P^{m,n}_{S}\lambda^{m}_{SD}
+P^{m,n}_{R}\lambda^{n}_{RD}\right),
\end{equation}
that is,
\begin{equation}
\left(1+P^{m,n}_{S}\lambda^{m}_{SR}\right)=\left(1+P^{m,n}_{S}\lambda^{m}_{SD}+P^{m,n}_{R}\lambda^{n}_{RD}\right).
\end{equation}
Together with $P^{m,n}=P^{m,n}_{S}+P^{m,n}_{R}$, we
obtain
%\begin{equation}
%\begin{split}
%&P^{m,n}_{S}=\frac{\lambda^{n}_{R,D}}{\lambda^{m}_{S,R}+\lambda^{n}_{R,D}-\lambda^{m}_{S,D}}P^{m,n}\\
%&P^{m,n}_{R}=\frac{\lambda^{m}_{S,R}-\lambda^{m}_{S,D}}{\lambda^{m}_{S,R}+\lambda^{n}_{R,D}-\lambda^{m}_{S,D}}P^{m,n}
%\end{split}
%\end{equation}

\begin{equation}
\left\{ {\begin{array}{lp{5mm}l}
   {P^{m,n}_{S}=\frac{\lambda^{n}_{RD}}{\lambda^{m}_{SR}+\lambda^{n}_{RD}-\lambda^{m}_{SD}}P^{m,n}},
   \\ {P^{m,n}_{R}=\frac{\lambda^{m}_{SR}-\lambda^{m}_{SD}}{\lambda^{m}_{SR}+\lambda^{n}_{RD}-\lambda^{m}_{SD}}P^{m,n}.}  \\
\end{array} } \right.
\end{equation}
%\begin{equation}
%\left\{ {\begin{array}{lp{5mm}l}
%   {P^{m,n}_{S}=\frac{\lambda^{n}_{R,D}}{\lambda^{m}_{S,R}+\lambda^{n}_{R,D}-\lambda^{m}_{S,D}}P^{m,n}\\
%   P^{m,n}_{R}=\frac{\lambda^{m}_{S,R}-\lambda^{m}_{S,D}}{\lambda^{m}_{S,R}+\lambda^{n}_{R,D}-\lambda^{m}_{S,D}}P^{m,n}.}  \\
%\end{array} } \right.
%\end{equation}
When the system works in the idle mode, we can easily get
\begin{equation}
\left\{ {\begin{array}{lp{5mm}l}
   {P^{m,n}_{S}=P^{m,n}}, \\ {P^{m,n}_{R}=0.}  \\
\end{array} } \right.
\end{equation}
Denote $\lambda^{m,n}$ as the equivalent channel gain given by
\begin{equation}  \label{eqv}
\lambda^{m,n}=\left \{\begin{array} {c@{\quad
\quad}l} \frac{\lambda^{m}_{SR}\lambda^{n}_{RD}}{\lambda^{m}_{SR}+\lambda^{n}_{RD}-\lambda^{m}_{SD}}, & \mbox{relaying\; mode},\\
\lambda^{m}_{S,D},& \mbox{idle \;mode}.
\end{array}
\right.
\end{equation}
By now, we can unify the rate as
\begin{equation}   \label{achrate}
R^{m,n}=\log_{2}\left(1+P^{m,n}\lambda^{m,n}\right).
\end{equation}

We define a subcarrier pairing parameter $t_{m,n}\in\{0,1\}$, which
takes $1$ if $\mbox{SP}(m,n)$ is selected, and $0$ otherwise. Then
the sum rate optimization problem can be formulated as
\begin{equation}   \label{ori0}
\begin{split}
&\underset{\{\mathbf{P},\mathbf{t}\}}\max\sum^{N}_{m=1}\sum^{N}_{n=1}t_{m,n}R^{m,n}, \\
s.t.&\quad\mathbf{C}1: \sum^{N}_{m=1}\sum^{N}_{n=1}t_{m,n}P^{m,n}\leq P_{t},   \;\mathbf{C}2:P^{m,n}\geq0 \;, \forall m,n, \\
&\quad\mathbf{C}3:\sum^{N}_{m=1}t_{m,n}=1, \forall n,\quad\quad\quad\mathbf{C}4:\sum^{N}_{n=1}t_{m,n}=1, \forall m,\\
%&\mathbf{C}5:t_{m,n}\in\{0,1\}, \forall m,n.
\end{split}
\end{equation}
where $P_{t}$ is the transmit power budget, $\mathbf{t}$ and
$\mathbf{P}$ are two $N\times N$ matrices with the $(m,n)$-th entry
$t_{m,n}$ and $P^{m,n}$ respectively. $\mathbf{C}3$ and $\mathbf{C}4$ correspond to the pairing constraint that each
subcarrier $m$ in listening phase only pairs with one subcarrier $n$ in relaying phase.

Since it is a mixed integer programming problem that is difficult to
solve, we relax the integer constraint of $t_{m,n}\in \{0,1\}$ as
$t_{m,n}\in [0,1],\, \forall m,n$ as
in~\cite{IEEEconf:35},~\cite{IEEEconf:32}. Denote
$S^{m,n}=t_{m,n}P^{m,n}$ as the actual power consumed on
$\mbox{SP}(m,n)$. Then the optimization problem becomes
\begin{equation}  \label{Rr}
\begin{split}
\underset{\{\mathbf{S},\mathbf{t}\}}\max\sum^{N}_{m=1}&\sum^{N}_{n=1}
t_{m,n}\frac{1}{2}\log_{2}\left(1+S^{m,n}\frac{\lambda^{m,n}}{t_{m,n}}\right),\\
s.t.\;\;&\mathbf{C}5: t_{m,n}\geq0,\forall
m,n,\\
&\mathbf{C}6: \sum^{N}_{m=1}\sum^{N}_{n=1}S^{m,n}\leq P_{t},\\
&\mathbf{C}7: S^{m,n}\geq0 \;, \forall
m,n,\;\mbox{and}\;\mathbf{C}3-\mathbf{C}4,
\end{split}
\end{equation}
where $\mathbf{S}=(S^{m,n})_{N\times N}$ is an $N\times N$ matrix.
Obviously the above objective function is concave with respect to
$(\mathbf{S},\mathbf{t})$. In the following, we will employ dual
method~\cite{IEEEconf:43,IEEEconf:33} to solve this optimization
problem.

In~\cite{IEEEconf:43}, the authors have shown that under a so-called
time-sharing condition, the duality gap of the optimization problem
is always zero, regardless of the convexity of the objective
function. Further, the authors show that the time-sharing condition
is always satisfied for practical multiuser spectrum optimization
problems in multi-carrier systems when the number of frequency
carriers goes to infinity. This suggests that we can solve the
problem by the dual method~\cite{IEEEconf:33}, which will provide an
upper bound for the original problem. More importantly, the method
can guarantee $t_{m,n}$ being integer-valued.

\subsection{Solution by the Dual Method}

Dualizing the constraints $\mathbf{C}4$ and $\mathbf{C}6$, we obtain the generated
Lagrange function as
\begin{equation} \label{Lag01}
\begin{split}
L(&\mathbf{S},\mathbf{t},\alpha,\mathbf{\beta})=\frac{1}{2}\sum^{N}_{m=1}\sum^{N}_{n=1}t_{m,n}\log_{2}\left(1+S^{m,n}\frac{\lambda^{m,n}}{t_{m,n}}\right)\\
&+\alpha\left(P_{t}-\sum^{N}_{m=1}\sum^{N}_{n=1}S^{m,n}\right)+\sum^{N}_{m=1}\beta_{m}\left(1-\sum^{N}_{n=1}t_{m,n}\right),
\end{split}
\end{equation}
where $\alpha\geq0$ and
$\boldsymbol{\beta}=(\beta_{1},\beta_{2},...,\beta_{N})\succeq0$ are
dual variables. Then the dual objective function and the dual problem are respectively
\begin{equation}   \label{Rd}
\begin{split}
g(\alpha,\boldsymbol{\beta})=\underset{\{\mathbf{S},\mathbf{t}\}}\max
L(\mathbf{s},\mathbf{t},\alpha,\boldsymbol{\beta}),\;\;\;s.t.\;\mathbf{C}3,\; \mathbf{C}5, \;\mathbf{C}6,
%&P^{m,n}>0  \forall m,n,\\
%&t_{m,n}\geq0 \forall m,n,\\
%&\sum^{N}_{n=1}t_{m,n}=1 \forall m
\end{split}
\end{equation}
and
\begin{equation}  \label{Rd10}
\underset{\{\alpha,\boldsymbol{\beta}\}}\min
g(\alpha,\boldsymbol{\beta})\quad s.t.\;\;\alpha\geq0,
\boldsymbol{\beta}\succeq0.
\end{equation}

Since a dual function is always optimized by first optimizing some
variables and then optimizing the remaining ones~\cite{IEEEconf:33}.
We first optimize $S^{m,n}$ with the assumption that $\alpha$ and $\beta_{m}$ are given.
Taking partial differentiation of $L$ with respect to $S^{m,n}$, we have
\begin{equation}
\frac{\partial L}{\partial
S^{m,n}}=\frac{t_{m,n}}{2}\frac{\frac{\lambda_{m,n}}{t_{m,n}}}{1+S^{m,n}\frac{\lambda_{m,n}}{t_{m,n}}}-\alpha=0,
\end{equation}
that is
\begin{equation}
\frac{1}{2\left(\frac{S^{m,n}}{t_{m,n}}+\frac{1}{\lambda_{m,n}}\right)}-\alpha=0.
\end{equation}
Together with constraint $S^{m,n}\geq0$, we obtain the optimal
solution
\begin{equation} \label{optp}
S_{*}^{m,n}=t_{m,n}\left[\frac{1}{2\alpha}-\frac{1}{\lambda_{m,n}}\right]^{+},
\end{equation}
where $[x]^{+}= \max\{0,x\}$. We find that $S_{*}^{m,n}$ is
associated with the subcarrier pairing parameter $t_{m,n}$. To find the optimal
solution for $t_{m,n}$, we first substitute~(\ref{optp}) into~(\ref{Lag01}) to obtain the updated Lagrange function
\begin{equation}  \label{Lag2}
\begin{split}
L(\mathbf{p},\mathbf{t},\alpha,\mathbf{\beta})=&\sum^{N}_{m=1}\sum^{N}_{n=1}\frac{t_{m,n}}{2}\log_{2}\left(1+\lambda^{m,n}\left[\frac{1}{2\alpha}-\frac{1}{\lambda_{m,n}}\right]^{+}\right)\\
&+\alpha\left(P_{t}-\sum^{N}_{m=1}\sum^{N}_{n=1}t_{m,n}\left[\frac{1}{2\alpha}-\frac{1}{\lambda_{m,n}}\right]^{+}\right)\\
&+\sum^{N}_{m=1}\beta_{m}\left(1-\sum^{N}_{n=1}t_{m,n}\right)\\
=&\sum^{N}_{m=1}\sum^{N}_{n=1}t_{m,n}T_{m,n}+\left(\alpha P_{t}+\sum^{N}_{m=1}\beta_{m}\right),
\end{split}
\end{equation}
where
\begin{equation}
\begin{split}
T_{m,n}=&\frac{1}{2}\log_{2}\left(1+\lambda^{m,n}\left[\frac{1}{2\alpha}-
\frac{1}{\lambda_{m,n}}\right]^{+}\right)\\
&-\alpha\left[\frac{1}{2\alpha}-\frac{1}{\lambda_{m,n}}\right]^{+}-\beta_{m}.
\end{split}
\end{equation}

Since both
$T_{m,n}$
and $\left(\alpha P_{t}+\sum^{N}_{m=1}\beta_{m}\right)$ are
independent of $t_{m,n}$, we obtain the optimal $t^{*}_{m,n}$ for
any $n$ as
\begin{equation}   \label{optt0}
t^{*}_{m,n}=\left \{\begin{array} {c@{\; :
\;}l} 1 & m=\arg\underset{m=1,...,N}\max T_{m,n},\\
0 & otherwise.
\end{array}
\right.
\end{equation}
Suppose that there is an subcarrier $m$ corresponding to two different $n$. Then it is
conflict to the constraint $\mathbf{C}4$, which is embedded in the Lagrangian. Therefore, $\sum^{N}_{m=1}t^{*}_{m,n}=1,\forall n$. Since both
$S_{*}^{m,n}$ and $t^{*}_{m,n}$ include the dual variables $\alpha$ and
$\beta_{m}$, we have to find values $\alpha$ and $\beta_{m}$ that
minimize $g(\alpha,\mathbf{\beta}_{m})$. Given $S_{(i)}^{m,n}$
and $t^{(i)}_{m,n}$ in the $i$-th iteration, the optimal values of dual variables can be iteratively achieved by the sub-gradient method~\cite{IEEEconf:34},
\begin{equation}   \label{opta}
\left\{ {\begin{array}{lp{5mm}l}
   {\alpha^{(i+1)}=\alpha^{(i)}-a^{(i)}\left(P_{t}-\sum^{N}_{m=1}\sum^{N}_{n=1}S_{(i)}^{m,n}\right)
   }, \\ {\beta^{(i+1)}_{m}=\beta^{(i)}_{m}-b^{(i)}\left(1-\sum^{N}_{n=1}t^{(i)}_{m,n}\right),\,\, m=1,...,N,}  \\
\end{array} } \right.
\end{equation}
in which $i$ is the iteration number, $a^{(i)}$ and $b^{(i)}$ are
step sizes designed properly. Within each iteration,  the subcarrier
pairing parameter and power allocation vectors  can be respectively
updated by~(\ref{optp}) and~(\ref{optt0}) with the updated $\alpha$
and $\beta_{m}$. Then the algorithm to find the optimal resource
allocation vectors can be designed as in Algorithm~1.
%
%
%in , corresponding to $S^{m,n}_{(i)}$ Eq.. Then
%the optimal power allocation and subcarrier pairing can be designed
%as in .

\vspace{-0.1cm}\hspace{-\parindent}\rule{\linewidth}{1pt}\vspace{-0.1cm}
\vspace{-0.25cm}{\bf Algorithm 1} The Optimal Resource Allocation Algorithm\\
\rule{\linewidth}{0.75pt}

\hspace{-\parindent}{\underline{Step 1}}: Set $i=1$, and initialize
$\alpha^{(i)}$, $\beta^{(i)}_{m}$, $\varepsilon$ and $\max_{iter}$,

\hspace{-\parindent}{\underline{Step 2}}: If $(i < \max_{iter})$,
$a^{(i)}=b^{(i)}=0.01/\sqrt{i}$,

%\hspace{-\parindent}{\underline{Step 3}}: Compute
%$\left(\frac{1}{2}\log_{2}\left(1+\lambda^{m,n}\left[\frac{1}{2\alpha}-\frac{1}
% {\lambda_{m,n}}\right]^{+}\right)-\alpha\left[\frac{1}{2\alpha}-\frac{1}{\lambda_{m,n}}\right]^{+}
% -\beta_{m}\right)$, $\forall m,n$ using $\alpha=\alpha^{(i)}$, $\beta=\beta^{(i)}_{m}$,

\hspace{-\parindent}{\underline{Step 3}}: Compute $t^{(i)}_{m,n}$ by
Eq.~(\ref{optt0}) using
 $\alpha=\alpha^{(i)}$ and\\
\indent\indent\indent \,$\beta_{m}=\beta^{(i)}_{m}$,

\hspace{-\parindent}{\underline{Step 4}}: Compute $S^{m,n}_{(i)}$ by
Eq.~(\ref{optp}) using $\alpha=\alpha^{(i)}$ and\\
\indent\indent\indent \,$t_{m,n}=t^{(i)}_{m,n}$,

\hspace{-\parindent}{\underline{Step 5}}: Compute $\alpha^{(i+1)}$,
$\beta^{(i+1)}_{m}$ by Eq.~(\ref{opta}) using $\alpha=\alpha^{(i)}$,\\
\indent\indent\indent \,$\beta_{m}=\beta^{(i)}_{m}$, $S^{m,n}=S^{m,n}_{(i)}$ and $t_{m,n}=t^{(i)}_{m,n}$,

\hspace{-\parindent}{\underline{Step 6}}: If
 $\frac{|\alpha^{(i+1)}-\alpha^{(i)}|}{|\alpha^{(i+1)}|}<\varepsilon$ and $\frac{||\beta^{(i+1)}_{m}-\beta^{(i)}_{m} ||}{||
\beta^{(i)}_{m}||}< \varepsilon$,\\
\indent\indent\indent \,exit and output $\alpha^{*}=\alpha^{(i+1)}$,
$\boldsymbol{\beta}^{*}=\boldsymbol{\beta}^{(i+1)}$,\\
\indent\indent\indent\;\!\!
$S_{*}^{m,n}=S^{m,n}_{(i)}, \;\;and\;\;
t^{*}_{m,n}=t^{(i)}_{m,n}$;\\
\indent\indent\indent \,otherwise set $i=i+1$ and go to Step $2$.
\\
%\hfill$\blacksquare$
\rule{\linewidth}{.75pt}

Denote the optimal values of the original problem~(\ref{ori0}), the
relaxed problem~(\ref{Rr}), and the relaxed dual problem~(\ref{Rd})
as $R_{o}$, $R_{r}$ and $R_{d}$ respectively. It is obviously
$R_{d}\geq R_{r}\geq R_{o}$. Because the optimal $t^{*}_{m,n}$
achieved by solving~(\ref{Rd}) and~(\ref{Rd10}) satisfy
$\mathbf{C}3$, $\mathbf{C}4$ and $t_{m,n}\in\{0,1\}$, $R_{d}$ is
also the dual optimum value for problem~(\ref{ori0}). In our
simulation, we find that the duality gap is asymptotically zero when
the number of subcarriers is reasonably large. Based on the analysis
and simulations of~\cite{IEEEconf:43},~\cite{IEEEconf:44} as well as
our paper, it can be concluded that $R_{d}\doteq R_{r}\doteq R_{o}$
for most of the practical cases.
%
%Thus, for practical OFDM systems, the duality gap is virtually zero. We can then conclude that for most practical OFDM systems.

If the subcarrier number is $N$, the total number of all possible
pairing configuration is $N!$. The complexity of computing the
achieved rate~(\ref{ori0}) is $N$ for a given subcarrier pairing
scheme. Thus, the complexity of exhaustive search is $O(N\cdot N!)$,
which is prohibitively high. However, within each iteration of
Algorithm~1, the complexity of the proposed algorithm is dominated
by the computation of~(\ref{Lag2}), which is $O(N^{2})$ in terms of
logarithm and multiplication operations. The complexity of computing
the optimal power allocation and the sum rate is $O(N)$. Therefore,
the total complexity for Algorithm~1 is $O(kN^{2})$, where $k$ is
the number of iterations. It is obvious that the complexity is
tractable.

The authors in ~\cite{IEEEconf:43,IEEEconf:44} showed that the
duality gap of the optimization problem is always zero when
time-sharing condition is satisfied, regardless of the convexity of
the objective function. They also showed that the time-sharing
condition will be satisfied if the optimal value of the optimization
problem is a concave function of the constraints. In our case, the
optimal subcarrier pairing may vary as the power constraint changes.
So the maximum sum rate as a function of the sum power constraint
may have discrete changes in the slope at the transition points
where the optimal subcarrier pairing scheme changes. The sudden jump
in the slope might make the optimization non-concave with the sum
power. But \cite{IEEEconf:44} also indicates analytically and
through simulations that the concavity will be asymptotically
satisfied as the number of subcarriers becomes large. This is
because that the amount of discrete slope change tends to decrease
with more subcarriers since the bandwidth affected by each change
becomes narrower. Therefore the curve is expected to be more concave
as the number of subcarriers increases. However, the
\cite{IEEEconf:44} as well as we can not rigorously prove this in
mathematics. In our simulations, we found that the concavity is
mostly satisfied when the number of subcarriers is reasonably large,
which is consistent to that observed in \cite{IEEEconf:44}. For
example, when N=2, we have observed that only about $0.8\%$ of the
possible channel conditions will result in the nonconcavity, and
when N=4, the probability turns to be $0.2\%$, the sum rate is
almost always concave in the sum power constraint when N=8. So it
can be concluded that duality gap is virtually zero for most of the
practical OFDM cases. This will also be verified by our simulation
results.

\subsection{Lloyd Algorithm Based Codebook Design}

%It is well known that adjusting power allocation among each
%transmitter node on each subcarrier can improve system performance
%remarkably.
If perfect CSI can be achieved at the source and relay, the resource
allocation vectors can be simply determined by Algorithm~1.
However, as stated earlier, due to limited resource of feedback link, the full knowledge of CSI available
at the transmit sides is difficult in OFDM systems. To solve this problem, We propose a limited feedback algorithm for power allocation
and subcarrier pairing in this subsection. In this
algorithm, the destination, which is assumed to have full forward CSI, selects a resource allocation vector from an elaborately designed codebook upon receiving the current CSI, and transmits
its index to the source and relay through a limited number of
feedback bits.
%calculating the optimal power allocation
%and subcarrier pairing, design a codebook of quantized power allocation and
%subcarrier pairing, then
%
%may not
%be available at the source/relay, and thus  PA vectors at the destination, a quantized vector is fed
%back to both the source and relay
%
% CSI a in a practical system,
%particularly in the systems using frequency division multiplexing,
%
%%because this
%%needs too much extra spectral resource.
%In this subsection, a  is presented to solve this problem.   selects a resource allocation vector index and transmits
%it to the source and relay over a limited rate feedback channel.
This technique employs a codebook of quantized power allocation and
subcarrier pairing designed offline and equipped on the source, relay and
destination. The codebook construction for limited-bit feedback can be linked to a vector quantization
problem. We use Lloyd algorithm~\cite{IEEEconf:19} to search for good
resource allocation codebooks based on sum rate criterion.
%using the Lloyd
%algorithm
%When using instantaneous rate optimized
%allocation,we
%by using Lloyd .
%Unfortunately, the implementation is complicated due to
%the fact that the transmitters must have knowledge of the forward
%channel. This assumption is often difficult to implement,
%particularly in systems using frequency division multiplexing.
%The scheme we proposed here can solve this problem effectively.

To design the limited feedback based codebook, we have to construct and iteratively use
\emph{the nearest neighbor rule} and \emph{the centroid condition}, which play crucial roles in the Lloyd algorithm. In the proposed algorithm, the \emph{the centroid condition} is designed to select the optimal codeword with maximum system rate in a given region, while the task of \emph{the nearest neighbor rule} is to determine the region in which the vectors are closest to the optimal codeword of this region. Notice that the optimization of finding the regions
and optimal resource allocation scheme is equivalent
to designing a vector quantizer with a modified distortion
measure~\cite{IEEEconf:19}.
%Different performance
%criteria may have different type of distortion function.
Taking the optimal rate performance as the design criterion, we use the
error distance function to measure the average distortion. Using \emph{the centroid condition} and \emph{the nearest neighbor rule}
iteratively, the error distance will decreases.
%Actually, when Lloyd algorithm is terminated, the destination has
%just searched over all codewords in the codebook and selects the
%best resource allocation vector with maximum rate. It just need to
%send back $b=\log_{2}B$ bits representing the codeword index to the
%source and relay nodes.

Suppose that the destination has perfect CSI $\mathbf{h}=(\mathbf{h}_{SD},\mathbf{h}_{SR},\mathbf{h}_{RD})$, where $\mathbf{h}_{SD}=\left(h^{1}_{SD},...,h^{N}_{SD}\right)$,
$\mathbf{h}_{SR}=\left(h^{1}_{SR},...,h^{N}_{SR}\right)$,
$\mathbf{h}_{RD}=\left(h^{1}_{RD},...,h^{N}_{RD}\right)$
%$\biggl\{\mathbf{h}_{i},i\in\{SD,SR,RD\}\biggr\}$
respectively denote the CSIs of SD, SR, and RD at a particular
period. Given $b$ bits of feedback, the space defined by all
possible sets of $\mathbf{h}$ is quantized into
$\emph{B}=2^{\emph{b}}$ regions.
%
%
%Then the CSI can be defined by a 3-tuple
%vector
%.
%Suppose that the destination has perfect CSI.
%%We quantize it at
%%first.
%If the destination node carries out  bits feedback, the CSI space
%will be quantized into $\emph{B}=2^{\emph{b}}$ regions.

In the sequel, we set codeword as
$\mathbf{c}=\left\{\left(P^{m,n}_{S},P^{m,n}_{R},t_{m,n}\right)|m,n=1,...,N\right\}$,
and
%define $\mathbf{c}^{*}$ as the optimal solution of Eq.~(\ref{ori}). Besides, we
denote $R(\mathbf{c}|\mathbf{h})$ as the end-to-end sum rate of a
given channel condition $\mathbf{h}$ and codeword $\mathbf{c}$. Then
\begin{equation} \label{alg1}
R(\mathbf{c}|\mathbf{h})=\frac{1}{2}\sum^{N}_{m=1}\sum^{N}_{n=1}t_{m,n}\log_{2}\left(1+P^{m,n}\lambda^{m,n}\right).
\end{equation}
%Given the capacity of the system as (6), the optimization issue
%becomes
%\begin{equation}
%\begin{split}
% \underset{A}\max \;C &(\mathbf{\alpha}_{s}, \mathbf{\alpha}_{r}|P_{tot}),\\
% s.t. &\mathbf{1}^{T}\alpha_{s}+\mathbf{1}^{T}\alpha_{r}=1,
% \\&\alpha_{s}\geq0, \alpha_{r}\geq0
% \end{split}
%\end{equation}

%Given the feedback bit $b$, the codebook size is $B=2^{b}$.
We first randomly generate the training channel condition set
$\mathbf{H}=\left\{\mathbf{h}_{l},l=1,...,M\right\}$ with $M\gg B$. Then we can easily obtain
the training code set
$\mathbf{T}=\{\mathbf{c}(\mathbf{h})|\mathbf{h}\in\mathbf{H}\}$, in which the $\mathbf{c}(\mathbf{h})$
denotes the optimal code achieved by Algorithm~1 for a given $\mathbf{h}\in\mathbf{H}$.
%
%
%training code set
%$\mathbf{T}=\{\mathbf{c}(\mathbf{h})|\mathbf{h}\in\mathbf{H}\}$.
%
%Given the
%%produce training codes $(T_{l}|l=0,1,...,M)$ by Algorithm~1.
%one can obtain an optimal code $\mathbf{c}(\mathbf{h})$ for each
%condition $\mathbf{h}\in\mathbf{H}$ by Algorithm~1. Then we obtain
%the training code set
%$\mathbf{T}=\{\mathbf{c}(\mathbf{h})|\mathbf{h}\in\mathbf{H}\}$.
The objective of Lloyd algorithm based codebook design is to randomly choose a
codebook
$\mathbb{C}=\{\mathbf{c}_{1},\mathbf{c}_{2},\ldots,\mathbf{c}_{B}\}$ of size $B$
from the training code set $\mathbf{T}$ and refine it. The error
distance function is defined as
\begin{equation}\label{error}
D(\mathbb{C})=E_{\mathbf{h}\in
\mathbf{H}}\left\{R(\mathbf{c}(\mathbf{h})|\mathbf{h})-\underset{0\leq
k \leq B}\max
 R(\mathbf{c}_{k}|\mathbf{h})\right\},
\end{equation}
where $E\{\cdot\}$ is the expectation of a random variable. Using this distortion function, the codebook design algorithm can be summarized as in Algorithm~2.
%\renewcommand{\description}[1]%
%{\hspace{\labelsep}\textsf{#1}}
%\newpage

\vspace{-0.1cm}\hspace{-\parindent}\rule{\linewidth}{1pt}\vspace{-0.1cm}
\vspace{-0.25cm}{\bf Algorithm 2} The Lloyd Algorithm Based Codebook Design\\
\rule{\linewidth}{0.75pt}

\hspace{-\parindent}{\underline{Step 1}}: Set $j=1$, $\varepsilon>0$, randomly generate the training code\\
\indent\indent\indent \,set and select the initial codebook\\
\indent\indent\indent \,$\mathbb{C}_{j}=\{\mathbf{c}^{j}_{1},\mathbf{c}^{j}_{2},\ldots,\mathbf{c}^{j}_{B}\}$
from $\mathbf{T}$, then calculate $D(\mathbb C_j)$\\
\indent\indent\indent \,by (\ref{error});

\hspace{-\parindent}{\underline{Step 2}}: Cluster the set of possible channel realization
vectors\\
\indent\indent\indent \,$\mathbf{H}$ into $B$ quantization regions with the $k$-th region\\
\indent\indent\indent \,$Q_{k}^j,\, k=1,...,B$, denoted as
\begin{equation*}
\begin{split}
Q^{j}_{k}=\biggl\{\mathbf{h}
%=\left(
%\mathbf{h}_{SD},\mathbf{h}_{SR},\mathbf{h}_{RD}\right)
|\left(
R(\mathbf{c}^{j}_{k}|\mathbf{h})\right) \geq \left(
R(\mathbf{c}^{j}_{l}|\mathbf{h})\right),\forall
l\in\{1,2,...,B\}\biggr\};
\end{split}
\end{equation*}

%\hspace{-\parindent}{\underline{Step 3}}: Compute
%$\left(\frac{1}{2}\log_{2}\left(1+\lambda^{m,n}\left[\frac{1}{2\alpha}-\frac{1}
% {\lambda_{m,n}}\right]^{+}\right)-\alpha\left[\frac{1}{2\alpha}-\frac{1}{\lambda_{m,n}}\right]^{+}
% -\beta_{m}\right)$, $\forall m,n$ using $\alpha=\alpha^{(i)}$, $\beta=\beta^{(i)}_{m}$,

\hspace{-\parindent}{\underline{Step 3}}: Using Eq.~(\ref{alg1}), generate a
new codebook $\mathbb{C}_{j+1}$ with\\
\indent\indent\indent \,the $k$-th codeword
% power allocation and
%subcarrier pairing vector
$\mathbf{c}^{j+1}_{k}$ defined as
\begin{equation*}
\begin{split}
\mathbf{c}^{j+1}_{k}=\arg \underset{\mathbf{c}\in \mathbf{T}}\max
E_{\mathbf{h}\in
Q^{j}_{k}}\left(R(\mathbf{c}|\mathbf{h})\right), \; k=1,2,...,B;
%\\&=\arg \underset{\mathbf{c}}\max \mathbf{E}_{T\in
%Q_{k}}\left(\sum^{N}_{m=1}\sum^{N}_{n=1}t_{m,n}\frac{1}{2}\log_{2}\left(1+P^{m,n}\lambda^{m,n}\right)\right).
\end{split}
\end{equation*}

\hspace{-\parindent}{\underline{Step 4}}: Calculate the average distortion $D(\mathbb{C}_{j+1})$
 by (\ref{error});

\hspace{-\parindent}{\underline{Step 5}}: If $D(\mathbb{C}_{j+1})<
D(\mathbb{C}_{j})+\varepsilon$ for some small $\varepsilon$, stop\\
\indent\indent\indent \,iteration and set the optimal
$\mathbb{C^{*}}=\mathbb{C}_{j+1}$;\\
\indent\indent\indent \,otherwise set $j=j+1$ and go back to Step $2$.
%
%\hspace{-\parindent}{\underline{Step 6}}: If
% $\frac{|\alpha^{(i+1)}-\alpha^{(i)}|}{|\alpha^{(i+1)}|}<\varepsilon$ and $\frac{||\beta^{(i+1)}_{m}-\beta^{(i)}_{m} ||}{||
%\beta^{(i)}_{m}||}< \varepsilon$,\\
%\indent \indent\indent\indent exit and output $\alpha^{*}=\alpha^{(i+1)}$,
%$\boldsymbol{\beta}^{*}=\boldsymbol{\beta}^{(i+1)}$
%%$\alpha=\alpha^{(i+1)}, \beta_{m}=\beta^{(i+1)}_{m},
%$S^{*}_{m,n}=S_{m,n}^{(i)}, \;\;and\;\;
%t^{*}_{m,n}=t^{(i)}_{m,n}$;\\
%\indent\indent\indent\indent otherwise set $i=i+1$ and go to Step $2$.
\\
%\hfill$\blacksquare$
\rule{\linewidth}{.75pt}

While the offline design of codebook seems to be computationally complex and time consuming, the real-time feed back process is quite simple.

\subsection{Feedback Scheme}
%As a matter of fact,
Upon receiving the instantaneous CSI $\mathbf{h}$, the destination searches over all the codewords in the designed codebook of size
$B$, and selects the $q$-th codeword provided with maximum sum rate,
i.e., $q=\arg \underset{q}\max
\left(R(\mathbf{c}_{q}|\mathbf{h})\right)$.
%
%
%${\mathbf{c}^{*}_{q}=\left(P^{m,n}_{S*},P^{m,n}_{R*},t^{*}_{m,n}\right)}$
%from the designed codebook whose size is $\emph{B}$, where
%$\mathbf{c}^{*}_{q}=\arg \underset{\mathbf{c}}\max
%                    \mathbf{E}_{\mathbf{h}\in
%                    Q_{q}}\left(R(\mathbf{c}|\mathbf{h})\right)$.
Afterward the destination sends back the index $q$ to both the source and relay through a noiseless feedback link. Since
the source and relay have been equipped with the same codebook copies, upon
receiving $q$, the source transmits with power $P^{m,n}_{S}$ and the relay with power $P^{m,n}_{R}$ indexed by $q$.

%=======================================================section4=====================================================================
\section{Optimal Resource Allocation for Enhanced DF Mode}\label{sec:4}
Depending on whether the relay is helpful, each subcarrier pairing
may work in either the relaying mode or the idle mode. For a
subcarrier pair working in the idle mode, the idle subcarrier in
the relaying phase is not utilized. We further allow the source to transmit
extra messages on those idle subcarriers in the relaying phase, which is called \emph{enhanced} DF mode in this paper.
%will
%certainly improve system rate. This relay  and we will analyze this enhanced DF scheme in this
%section.

\subsection{Formulation of the Optimization Problem}
Similarly, the achieved rate of the \emph{enhanced} DF mode is given at the top of the next page.
%\begin{equation}
%R^{m,n}=\left \{\begin{array} {c@{
%\quad}l} \frac{1}{2}\left\{\left(1+P^{m,n}_{S,(1)}\lambda^{m}_{S,D}\right)+\log_{2}\left(1+P^{m,n}_{S,(2)}\lambda^{n}_{S,D}\right)\right\}, & Modified\; idle\; mode,\\
%\frac{1}{2}\min\{\log_{2}\left(1+P^{m,n}_{S,(1)}\lambda^{m}_{S,R}\right),
%\log_{2}\left(1+P^{m,n}_{S,(1)}\lambda^{m}_{S,D}+P^{m,n}_{R}\lambda^{n}_{R,D}\right)\},& Relaying \;mode.
%\end{array}
%\right.
%\end{equation}
%\begin{displaymath}
%R^{m,n}=\begin{cases}
%\;\frac{1}{2}\left\{\log_{2}\left(1+P^{m,n}_{S,1}\lambda^{m}_{SD}\right)+\log_{2}\left(1+P^{m,n}_{S,2}\lambda^{n}_{SD}\right)\right\},\\ \;\quad\quad\quad\quad\qquad\qquad\qquad\qquad\qquad\qquad \mbox{idle mode},\\
%\;\frac{1}{2}\min\bigg\{\log_{2}\left(1+P^{m,n}_{S,1}\lambda^{m}_{SR}\right),\\
%  \qquad\qquad\qquad\log_{2}\left(1+P^{m,n}_{S,1}\lambda^{m}_{SD}+P^{m,n}_{R}\lambda^{n}_{RD}\right)\bigg\},\\
%  \;\qquad\qquad\qquad\qquad\qquad\qquad\qquad\mbox{relaying \;mode}.
%\end{cases}
%\end{displaymath}
%
\begin{figure*}[!t]
\begin{displaymath}\label{top}
R^{m,n}=\begin{cases}
\;\frac{1}{2}\left\{\log_{2}\left(1+P^{m,n}_{S,1}\lambda^{m}_{SD}\right)+
\log_{2}\left(1+P^{m,n}_{S,2}\lambda^{n}_{SD}\right)\right\},& \mbox{idle mode},\\
\;\frac{1}{2}\min\bigg\{\log_{2}\left(1+P^{m,n}_{S,1}\lambda^{m}_{SR}\right),
\log_{2}\left(1+P^{m,n}_{S,1}\lambda^{m}_{SD}+P^{m,n}_{R}\lambda^{n}_{RD}\right)\bigg\},&\mbox{relaying \;mode}.
\end{cases}
\end{displaymath}
\rule{\linewidth}{0.5pt}
\end{figure*}
%
%\begin{equation}
%\begin{split}
%R=&\frac{1}{2}\sum_{(m,n)\in
%S_{s}}\log_{2}\left(1+P^{m,n}_{S,(1)}\lambda^{m}_{S,D}\right)+\log_{2}\left(1+P^{m,n}_{S,(2)}\lambda^{n}_{S,D}\right)\\
%&+\frac{1}{2}\sum_{(m,n)\in
%S_{r}}\min\left\{\log_{2}\left(1+P^{m,n}_{S,(1)}\lambda^{m}_{S,R}\right),\log_{2}\left(1+P^{m,n}_{S,(1)}\lambda^{m}_{S,D}+P^{m,n}_{R}\lambda^{n}_{R,D}\right)\right\}
%\end{split}
%\end{equation}
$P^{m,n}_{S,1}$, $P^{m,n}_{S,2}$ and $P^{m,n}_{R}$
respectively denote the source power in the listening phase, the source
power in the relaying phase and the relay power in the relaying phase.
%Then
%the transmit power on $\mbox{SP}(m,n)$ is
%$P^{m,n}=P^{m,n}_{S,1}+P^{m,n}_{S,2}+P^{m,n}_{R}$.
\setcounter{equation}{23}
Because the condition to activate the relay depends not only on the channel gains but also on the power allocation, we define an indicator $\rho_{m,n}\in\{0,1\}$ to show the status of $\mbox{SP}(m,n)$ at relay, i.e., the relay is used for $\mbox{SP}(m,n)$ if $\rho_{m,n}=1$, otherwise, it is not used. Let
the equivalent channel gain
$\lambda^{m,n}_{1}=\frac{\lambda^{m}_{SR}\lambda^{n}_{RD}}{\lambda^{m}_{SR}
+\lambda^{n}_{RD}-\lambda^{m}_{SD}}$, and let
$P^{m,n}_{SR}=P^{m,n}_{S,1}+P^{m,n}_{R}$. Then the optimization
problem based on the sum rate of the \emph{enhanced} DF mode can be formulated
as
\begin{equation}
\begin{split}
&\underset{\{\mathbf{P},\mathbf{t},\mathbf{\rho}\}}\max\frac{1}{2}\sum^{N}_{m=1}\sum^{N}_{n=1}t_{m,n}\bigg\{\rho_{m,n}\log_{2}\left(1+P^{m,n}_{SR}\lambda^{m,n}_{1}\right)
\\&\quad\quad\quad\quad+(1-\rho_{m,n})\log_{2}\left(1+P^{m,n}_{S,1}\lambda^{m}_{SD}\right)\\
&\quad\quad\quad\quad+(1-\rho_{m,n})\log_{2}\left(1+P^{m,n}_{S,2}\lambda^{n}_{SD}\right)
\bigg\},\\
s.t.\quad %&\mathbf{D}1: t_{m,n}\in\{0,1\},\forall m\;n,\\
&\mathbf{D}1: \sum^{N}_{m=1}t_{m,n}=1, \forall n,\;\sum^{N}_{n=1}t_{m,n}=1, \forall m,\\
%&\mathbf{D}3: \rho_{m,n}\in\{0,1\},\\
&\mathbf{D}2: \sum^{N}_{m=1}\sum^{N}_{n=1}t_{m,n}\bigg\{(1-\rho_{m,n})(P^{m,n}_{S,1}+P^{m,n}_{S,2})\\
&\quad\quad\quad\quad\quad\quad\quad\quad+\rho_{m,n}P^{m,n}_{SR}\bigg\}\leq P_{t},\\
&\mathbf{D}3: P^{m,n}_{S,1},P^{m,n}_{S,2},P^{m,n}_{SR}\geq 0, \forall\;m \;n,\\
%&\mathbf{D}6: 0\leq \rho_{m,n} \leq1
\end{split}
\end{equation}
where
$\mathbf{p}=\left(P^{m,n}_{SR},P^{m,n}_{S,1},P^{m,n}_{S,2}\right)
\in (\mathbf{R}^3)^{N\times N}$, $\mathbf{t}=\left(t_{m,n}\right)\in
\mathbf{R}^{N\times N}$ and
$\mathbf{\mathrm{\rho}}=\left(\rho_{m,n}\right)\in
\mathbf{R}^{N\times N}$.

Similarly, we make a continuous
relaxation to the optimization problem and obtain a standard convex problem. Moreover, we respectively denote $S_{SR}^{m,n}=t_{m,n}\rho_{m,n}P_{SR}^{m,n}$,
$S_{S,1}^{m,n}=t_{m,n}(1-\rho_{m,n})P_{S,1}^{m,n}$ and
$S_{S,2}^{m,n}=t_{m,n}(1-\rho_{m,n})P_{S,2}^{m,n}$ as the
actual power consumption at the source and the relay in the two phases. Then the relaxed optimization problem is formulated as
\begin{equation}
\begin{split}
\underset{\{\mathbf{s},\mathbf{t},\mathbf{\rho}\}}\max\frac{1}{2}&\sum^{N}_{m=1}\sum^{N}_{n=1}t_{m,n}\bigg\{\rho_{m,n}\log_{2}\left(1+S^{m,n}_{SR}\frac{\lambda^{m,n}_{1}}{t_{m,n}\rho_{m,n}}\right)
\\&+(1-\rho_{m,n})\log_{2}\left(1+S_{S,1}^{m,n}\frac{\lambda^{m}_{SD}}{t_{m,n}(1-\rho_{m,n})}\right)
\\&+(1-\rho_{m,n})\log_{2}\left(1+S_{S,2}^{m,n}\frac{\lambda^{n}_{SD}}{t_{m,n}(1-\rho_{m,n})}\right)\bigg\},\\
s.t.\quad &\mathbf{D}4: t_{m,n}\geq0,\forall m\;n,\;\;\mathbf{D}5: \rho_{m,n}\geq0,\forall m\;n,\\
&\mathbf{D}6:
\sum^{N}_{m=1}\sum^{N}_{n=1}\left(S_{SR}^{m,n}+S_{S,1}^{m,n}+S_{S,2}^{m,n}\right)= P_{t},\\
&\mathbf{D}7: S^{m,n}_{S,1},S^{m,n}_{S,2},S^{m,n}_{SR}\geq 0,
\forall\;m \;n, \;and\; \mathbf{D}1.
\end{split}
\end{equation}

\subsection{Dual Solution of the Relaxed Problem}
Dualizing the constraints $\mathbf{D}1$ and $\mathbf{D}6$, we obtain the
Lagrangian
\begin{equation}
\begin{split}
L(\mathbf{s},\mathbf{t},\mathbf{\rho},\mathbf{\alpha},\mathbf{\beta})&=\sum^{N}_{m=1}\sum^{N}_{n=1}
\frac{t_{m,n}}{2}\bigg\{\rho_{m,n}\log_{2}\left(1+S^{m,n}_{SR}\frac{\lambda^{m,n}_{1}}{t_{m,n}\rho_{m,n}}\right)
\\&+(1-\rho_{m,n})\log_{2}\left(1+S^{m,n}_{S,1}\frac{\lambda^{m}_{SD}}{t_{m,n}(1-\rho_{m,n})}\right)
\\&+(1-\rho_{m,n})\log_{2}\left(1+S^{m,n}_{S,2}\frac{\lambda^{n}_{SD}}{t_{m,n}(1-\rho_{m,n})}\right)\bigg\}\\
&+\alpha\left(P_{t}-\sum^{N}_{m=1}\sum^{N}_{n=1}\sum^{2}_{i=1}(S^{m,n}_{S,i}+S^{m,n}_{SR})\right)\\
&+\sum^{N}_{n=1}\beta_{n}\left(1-\sum^{N}_{m=1}t_{m,n}\right),
\end{split}
\end{equation}
where $\alpha$ and $\beta_{n}$ are dual variables as before. Then
the dual objective function and the dual problem can be respectively
expressed as
\begin{equation}
g(\alpha,\boldsymbol{\beta})=\underset{\{\mathbf{s},\mathbf{t},\boldsymbol{\rho}
\}}\max
L(\mathbf{s},\mathbf{t},\boldsymbol{\rho},\mathbf{\alpha},\boldsymbol{\beta}),
\;\; s.t. \;\mathbf{D}1,\; \mathbf{D}4-\mathbf{D}6,
\end{equation}
and
\begin{equation}
\underset{\{\mathbf{\alpha},\boldsymbol{\beta}\}}\min
g(\alpha,\boldsymbol{\beta}) \;\; s.t.\quad\alpha\geq0,
\boldsymbol{\beta}\succeq0.
\end{equation}
Taking derivatives of $L$ with respect to $S^{m,n}_{SR}$, $S^{m,n}_{S,1}$ and $S^{m,n}_{S,2}$, we obtain the
optimal solutions
%Then we have
%\begin{equation}
%\frac{\partial L}{\partial
%S^{m,n}_{SR}}=\frac{t_{m,n}\rho_{m,n}}{2}\frac{\lambda^{m,n}_{1}}{t_{m,n}\rho_{m,n}+S^{m,n}_{SR}\lambda^{m,n}_{1}}-\alpha=0,
%\end{equation}
%together with the constraint $S^{m,n}_{SR}\geq0$, we can obtain
\begin{equation}
S_{SR*}^{m,n}=t_{m,n}\rho_{m,n}\left[\frac{1}{2\alpha}-\frac{1}{\lambda^{m,n}_{1}}\right]^{+},
\end{equation}
%Then we take derivatives of $L$ with respect to $S^{m,n}_{S,1}$ and
%have
%\begin{equation}
%\frac{\partial L}{\partial
%S^{m,n}_{S,1}}=1+S^{m,n}_{S,1}\frac{\lambda^{m}_{SD}}{t_{m,n}(1-\rho_{m,n})}-\frac{\lambda^{m}_{SD}}{2\alpha}=0,
%\end{equation}
%together with the constraint $S^{m,n}_{S,1}\geq0$, we obtain the
%optimal solution
\begin{equation}
S_{S,1*}^{m,n}=t_{m,n}(1-\rho_{m,n})\left[\frac{1}{2\alpha}-\frac{1}{\lambda^{m}_{SD}}\right]^{+},
\end{equation}
%Similarly we obtain the optimal solution of $S^{m,n}_{S,2}$ as
and
\begin{equation}
S_{S,2*}^{m,n}=t_{m,n}(1-\rho_{m,n})\left[\frac{1}{2\alpha}-\frac{1}{\lambda^{n}_{SD}}\right]^{+}.
\end{equation}
Denote $R_{m,n}^{R}$ as the rate contribution of $\mbox{SP}(m,n)$ to the Lagrangian in the relaying mode, and $R_{m,n}^{I}$ in the idle mode. Then we have
\begin{equation}  \label{rr}
\begin{split}
&R_{m,n}^{R}=\\
&\quad\frac{1}{2}\log
\left(1+\lambda^{m,n}_{1}\left[\frac{1}{2\alpha}-\frac{1}{\lambda^{m,n}_{1}}\right]^{+}\right)
-\alpha\left[\frac{1}{2\alpha}-\frac{1}{\lambda^{m,n}_{1}}\right]^{+},
\end{split}
\end{equation}
\begin{equation}  \label{ri}
\begin{split}
&R_{m,n}^{I}=\\
&\quad\frac{1}{2}\log
\left(1+\lambda^{m}_{SD}\left[\frac{1}{2\alpha}-\frac{1}{\lambda^{m}_{SD}}\right]^{+}\right)
-\alpha\left[\frac{1}{2\alpha}-\frac{1}{\lambda^{m}_{SD}}\right]^{+}\\
&\;\;+\frac{1}{2}\log
\left(1+\lambda^{n}_{SD}\left[\frac{1}{2\alpha}-\frac{1}{\lambda^{n}_{SD}}\right]^{+}\right)
-\alpha\left[\frac{1}{2\alpha}-\frac{1}{\lambda^{n}_{SD}}\right]^{+}.
\end{split}
\end{equation}
Easily we obtain the optimal indictor as
\begin{equation}    \label{optr}
\rho^{*}_{m,n}=\left \{\begin{array} {c@{\;
\;}l} 1,\quad & \mbox{when} \quad R_{m,n}^{R}>R_{m,n}^{I},\\
0,\quad & \mbox{otherwise}.
\end{array}
\right.
\end{equation}
Denote
$R_{m,n}^{*}=\rho^{*}_{m,n}R_{m,n}^{R}+(1-\rho^{*}_{m,n})R_{m,n}^{I}-\beta_{n}$ for briefness,
we obtain the optimal subcarrier pairing parameter as
\begin{equation}
t^{*}_{m,n}=\left \{\begin{array} {c@{\;
\;}l} 1, \quad& m=\underset{m=1,...,N}{\arg\max} R_{m,n}^{*}, \\
0, \quad& \mbox{otherwise},
\end{array}
\right.  \forall n.
\end{equation}
We similarly
update the Lagrange multipliers $\alpha$ and $\boldsymbol{\beta}$ by
subgradient method as
\begin{equation}
\left\{ {\begin{array}{lp{5mm}l}
   {\alpha^{(i+1)}=\alpha^{(i)}-a^{(i)}\bigg\{P_{t}-}\\
   {\quad\quad\quad\quad\quad\quad\sum^{N}_{m=1}\sum^{N}_{n=1}\left(S_{SR}^{m,n}+S_{S,1}^{m,n}+S_{S,2}^{m,n}\right)\bigg\}
   }, \\ {\beta^{(i+1)}_{m}=\beta^{(i)}_{m}-b^{(i)}\left(1-\sum^{N}_{n=1}t^{(i)}_{m,n}\right),\quad m=1,...,N.}  \\
\end{array} } \right.
\end{equation}
With the updated $\alpha$ and $\beta_{m}$ in each iteration, we can update
the subcarrier pairing $t^{*}_{m,n}$, the power allocation
vectors $\left(S_{SR*}^{m,n},S_{S,1*}^{m,n},S_{S,2*}^{m,n}\right)$
as well as the indicator $\rho^{*}_{m,n}$ by Algorithm~1.
%
%
%By now, we have found the optimal subcarrier pairing vector
% as well as the power allocation vector
%$\left(P_{SR*}^{m,n},P_{S,1*}^{m,n},P_{S,2*}^{m,n}\right)$
%corresponding to
%$\left(S_{SR*}^{m,n},S_{S,1*}^{m,n},S_{S,2*}^{m,n}\right)$
%respectively, the mode selection vector $\rho^{*}_{m,n}$ and the
%dual variables $\alpha$ and $\mathbf{\beta}_{m}$. Then we can update
%the subcarrier pairing and power allocation vectors with the renewed
%$\alpha$ and $\beta_{m}$ in each iteration as in Algorithm~1.
Notice that the iteration procedure in Algorithm~1 should be modified in
some places. For example, before computing $t^{(i)}_{m,n}$ in the Algorithm~1, we have to figure out $R_{m,n}^{R}$,
$R_{m,n}^{I}$ and $\rho^{*}_{m,n}$ by~(\ref{rr}),~(\ref{ri})
and~(\ref{optr}) respectively. Similarly, we can use Algorithm~2 to design codebook for limited feedback.

%Besides, in Algorithm~2, the limited feedback based codeword
%designed offline with the Lloyd Algorithm has the codeword of the
%form
%$\left(P^{m,n}_{SR},P^{m,n}_{S,1},P^{m,n}_{S,2},t_{m,n},\rho_{m,n}\right)$.
%Similarly, upon receiving the current CSI, the destination searches over all
%codewords in the designed codebook, select the one with maximum
%associated rate, and broadcasts the codeword index to both the
%source and relay.
%It is easy to find that the selection process is
%quite simple although the offline design of codebook may seem to be
%computational complex.

\subsection{Resource Allocation Under Individual Power Constraints}

In this subsection, we investigate the resource allocation under individual power constraints
for the source and the relay.
For the individual power constraints, the sum powers at the source and the relay have separate constraints, which can be expressed as:
\begin{equation}
\sum^{N}_{m=1}P_{S}^{m,n}\leq P_{S},\;\sum^{N}_{n=1}P_{R}^{m,n}\leq P_{R},
\end{equation}
where $P_{S}$ and $P_{R}$ denote the source power constraint and the relay power constraint respectively.
For a given subcarrier pairing~$\mbox{SP}(m,n)$, the mode selection criterion~\cite{IEEEconf:31},~\cite{IEEEconf:60} is expressed as
\begin{equation}
\mbox{Relaying \;mode}:\;\lambda_{SR}^{m}P_{S}^{m,n}\geq\lambda_{SD}^{m}P_{S}^{m,n}
+\lambda_{RD}^{n}P_{R}^{m,n}.
\end{equation}
Then we can similarly obtain a Lagrangian
\begin{equation}
\begin{split}
L=&\sum^{N}_{m=1}\sum^{N}_{n=1}R^{m,n}\\
&+\mu_{S}\left(P_{S}-\sum^{N}_{m=1}P_{S}^{m,n}\right)
+\mu_{R}\left(P_{R}-\sum^{N}_{n=1}P_{R}^{m,n}\right)\\
&+\underset{\{m,n\}\in S_{R}}\sum\rho_{m,n}\left(\lambda_{SR}^{m}P_{S}^{m,n}-
\lambda_{SD}^{m}P_{S}^{m,n}-\lambda_{RD}^{n}P_{R}^{m,n}\right),
\end{split}
\end{equation}
where $S_{R}$ is SP set of relaying mode. The Lagrange coefficients
$\mu_{S},\,\mu_{R}\geq0$ are chosen such that the individual power
constraints are satisfied. Lagrange multiplier $\rho_{m,n}\geq0$
corresponds to the mode selection criterion. For almost all of the
subcarrier pairs belonging to relaying mode, the authors
in~\cite{IEEEconf:60} conclude that the selection criterion will be
satisfied when
\begin{equation}  \label{decodecons}
\lambda_{SR}^{m}P_{S}^{m,n}=\lambda_{SD}^{m}P_{S}^{m,n}+\lambda_{RD}^{n}P_{R}^{m,n},
\end{equation}
with a possible exception pair satisfying $\gamma_{RD}^{n}/\gamma_{SD}^{m}=\lambda_{R}/\lambda_{S}$. However, usually there will be at most one subcarrier pair
in this set. Fortunately, we find that the exception $\mbox{SP}(m,n)$ have the same contribution and cost to the Lagrangian in the model, which are respectively $\frac{1}{2}\log\left(\frac{\lambda_{SD}^{m}}{2\mu_{S}}\right)$ and $\mu_{S}\left(\frac{1}{2\mu_{S}}-\frac{1}{\lambda_{SD}^{m}}\right)$, no matter it is classified into relaying mode or idle mode. So we assign it to relaying mode thereafter.

For relaying mode, (\ref{decodecons}) implies $P_{S}^{m,n}=\frac{\lambda_{RD}^{n}}{\lambda_{SR}^{m}-\lambda_{SD}^{m}}P_{R}^{m,n}$.
Then $P_{S}^{m,n}$ and $P_{R}^{m,n}$ will be zero or positive
simultaneously. Thus in the relaying mode, we can first allocate
total power of $\mbox{SP}(m,n)$ and then obtain the corresponding
$P_{S}^{m,n}$ and $P_{R}^{m,n}$. Let
\begin{equation}   \label{fen}
\left \{\begin{array} {c@{\;
\;}l} P_{S}^{m,n}=\frac{\lambda^{n}_{RD}}{\lambda^{m}_{SR}+\lambda^{n}_{RD}-\lambda^{m}_{SD}}P^{m,n}, \\
P_{R}^{m,n}=\frac{\lambda^{m}_{SR}-\lambda^{m}_{SD}}{\lambda^{m}_{SR}+\lambda^{n}_{RD}-
\lambda^{m}_{SD}}P^{m,n},
\end{array}
\right.
\end{equation}
in the relaying mode, and
\begin{equation}
\left\{ {\begin{array}{lp{5mm}l}
   {P^{m,n}_{S}=P^{m,n}}, \\ {P^{m,n}_{R}=0.}  \\
\end{array} } \right.
\end{equation}
in the idle mode.
Denote the equivalent channel gain of $\mbox{SP}(m,n)$ by
%\begin{equation} \label{dengxiao}
%\overline{\gamma}^{m,n}=\frac{\gamma^{m}_{S,R}\gamma^{n}_{R,D}}{\gamma^{m}_{S,R}
%+\gamma^{n}_{R,D}-\gamma^{m}_{S,D}}.
%\end{equation}
\begin{equation}  \label{dengxiao}
\overline{\lambda}^{m,n}=\left \{\begin{array} {c@{\quad
\quad}l} \frac{\lambda^{m}_{SR}\lambda^{n}_{RD}}{\lambda^{m}_{SR}+\lambda^{n}_{RD}-\lambda^{m}_{SD}}, & \mbox{relaying\; mode},\\
\lambda^{m}_{SD},& \mbox{idle \;mode}.
\end{array}
\right.
\end{equation}
We can also use an unified rate expression to demonstrate the original optimization as
\begin{equation}
R_{m,n}=\frac{1}{2}\log(1+\overline{\lambda}^{m,n}P^{m,n}).
\end{equation}
%
%
%\begin{displaymath}
%\begin{cases}
%\;P_{S}^{m,n}=\frac{\gamma^{n}_{R,D}}{\gamma^{m}_{S,R}+\gamma^{n}_{R,D}-\gamma^{m}_{S,D}}P^{m,n}, \\
%\;P_{R}^{m,n}=\frac{\gamma^{m}_{S,R}-\gamma^{m}_{S,D}}{\gamma^{m}_{S,R}+\gamma^{n}_{R,D}-\gamma^{m}_{S,D}}P^{m,n},
%\end{cases}
%\end{displaymath}
%
%\begin{equation}
%P_{S}^{m,n}=\left \{\begin{array} {c@{\quad
%\quad}l} \frac{\lambda^{n}_{R,D}}{\lambda^{m}_{S,R}+\lambda^{n}_{R,D}-\lambda^{m}_{S,D}}P^{m,n}, & (m,n)\in S_{R}\cup S_{I},\\
%P^{m,n},& (m,n)\in S_{S}.
%\end{array}
%\right.
%\end{equation}
%\begin{displaymath}
%P_{R}^{m,n}=\begin{cases}
%\;\frac{\gamma^{m}_{S,R}-\gamma^{m}_{S,D}}{\gamma^{m}_{S,R}+\gamma^{n}_{R,D}-\gamma^{m}_{S,D}}P^{m,n}, & (m,n)\in S_{R}\cup S_{I},\\
%\;0,& (m,n)\in S_{S}.
%\end{cases}
%\end{displaymath}
%\begin{displaymath}
%
%\begin{cases}
%\;, & (m,n)\in S_{R}\cup S_{I},\\
%\;\gamma^{m}_{S,D},& (m,n)\in S_{S},
%\end{cases}
%\end{displaymath}
The unified rate helps simplifying
the optimization in the same way.
Let
\begin{equation}
\begin{split}
\mathbf{R}=\sum^{N}_{m=1}&\sum^{N}_{n=1}\frac{t_{m,n}}{2}\bigg\{\rho_{m,n}\log_{2}\left(1+
\frac{P^{m,n}_{1}\overline{\lambda}^{m,n}}{t_{m,n}\rho_{m,n}}\right)
\\&\;\;+(1-\rho_{m,n})\biggl[\log_{2}\left(1+\frac{P_{2}^{m,n}\lambda^{m}_{SD}}{t_{m,n}(1-\rho_{m,n})}\right)
\\&\qquad\qquad\;\;+\log_{2}\left(1+\frac{P_{3}^{m,n}\lambda^{n}_{SD}}{t_{m,n}(1-\rho_{m,n})}\right)\biggr]\bigg\},
\end{split}
\end{equation}
where $P^{m,n}_{1}$ is the sum power of $\mbox{SP}(m,n)$ in the relaying mode, which can be obtained from~(\ref{fen}). $P_{2}^{m,n}$ and $P_{3}^{m,n}$ are respectively the powers used by the
direct-link of $\mbox{SP}(m,n)$ in the listening and relaying phases.
Then the
sum rate optimization is formulated as
\begin{equation}  \label{ori}
\begin{split}
&\underset{\{\mathbf{S},\mathbf{t},\mathbf{\boldsymbol{\rho}}\}}\max \;\; \;\mathbf{R},\\
s.t.\; &\mathbf{E}1: \sum^{N}_{m=1}t_{m,n}=1, \forall n,\;\;\;\;\mathbf{E}2:\sum^{N}_{n=1}t_{m,n}=1, \forall m,\\
&\mathbf{E}3:\rho_{m,n}\in\{0,1\},\quad\quad\;\;\,\mathbf{E}4:t_{m,n}\in\{0,1\},\\
&\mathbf{E}5:\sum^{N}_{m=1}\sum^{N}_{n=1}(\eta^{m,n}_{S}P_{1}^{m,n} + P_{2}^{m,n} + P_{3}^{m,n})\leq P_{S},\\
&\mathbf{E}6:\sum^{N}_{m=1}\sum^{N}_{n=1}\eta^{m,n}_{R}P_{1}^{m,n}\leq P_{R},\;\;
\mathbf{E}7: P^{m,n}_{j}\geq 0,\forall j.
%
%\mathbf{D}4: t_{m,n}\geq0,\forall m\;n,\\
%&\mathbf{D}5: \rho_{m,n}\geq0,\forall m\;n,\\
%&\mathbf{D}6:
%\sum^{N}_{m=1}\sum^{N}_{n=1}\left(S_{1}^{m,n}+S_{2}^{m}+S_{3}^{n}\right)= P_{t},\\
%&\mathbf{D}7: S^{m,n}_{1},S^{m}_{2},S^{n}_{3}\geq 0,
% \;and\; \mathbf{D}1.
\end{split}
\end{equation}
Let $\mathbf{P}=\left(P^{m,n}_{1},P^{m,n}_{2},P^{m,n}_{3}\right)\in \mathbb{R}^{N\times N\times 3}$,
$\mathbf{t}=\left(t_{m,n}\right)\in
\mathbb{R}^{N\times N}$ and
$\mathbf{\boldsymbol{\rho}}=\left(\rho_{m,n}\right)\in
\mathbb{R}^{N\times N}$. Denote
\begin{equation}
\eta^{m,n}_{S}=\left \{\begin{array} {c@{\quad
\quad}l} \frac{\lambda^{n}_{R,D}}{\lambda^{m}_{SR}+\lambda^{n}_{RD}-\lambda^{m}_{SD}}, & \mbox{relaying\; mode},\\
1,& \mbox{idle \;mode},
\end{array}
\right.
\end{equation}
and
\begin{equation}
\eta^{m,n}_{R}=\left \{\begin{array} {c@{\quad
\quad}l} \frac{\lambda^{m}_{SR}-\lambda^{m}_{SD}}{\lambda^{m}_{SR}+\lambda^{n}_{RD}-\lambda^{m}_{SD}}, & \mbox{relaying\; mode},\\
0,& \mbox{idle \;mode}.
\end{array}
\right.
\end{equation}
We dualize the constraints $\mathbf{E}1$, $\mathbf{E}5$, $\mathbf{E}6$ and~(\ref{decodecons}). Then the generated Lagrange function is
\begin{equation} \label{Lag1}
\begin{split}
L(\mathbf{P},\mathbf{t},&\lambda_{S},\lambda_{R},\boldsymbol{\beta})=\mathbf{R}+\sum^{N}_{n=1}\beta_{n}\left(1-\sum^{N}_{m=1}t_{m,n}\right)\\
&+\mu_{S}\left(P_{S}-\sum^{N}_{m=1}\sum^{N}_{n=1}(\eta^{m,n}_{S}P^{m,n}_{1}-P^{m,n}_{2}-P^{m,n}_{3})\right)\\
&+\mu_{R}\left(P_{R}-\sum^{N}_{m=1}\sum^{N}_{n=1}\eta^{m,n}_{R}P^{m,n}_{1}\right),
%&+\underset{\{m,n\}\in S_{R}}\sum\rho_{m,n}\left(\gamma_{SR}^{m}P_{S}^{m,n}-\gamma_{SD}^{m}P_{S}^{m,n}-\gamma_{RD}^{n}P_{R}^{m,n}\right),
\end{split}
\end{equation}
%\begin{equation} \label{Lag1}
%\begin{split}
%L&(\mathbf{P},\mathbf{t},\lambda_{S},\lambda_{R},\boldsymbol{\beta})=\mathbf{R}+\sum^{N}_{n=1}\beta_{n}\left(1-\sum^{N}_{m=1}t_{m,n}\right)\\
%&\;+\lambda_{S}\left(P_{S}-\sum^{N}_{m=1}\sum^{N}_{n=1}(\eta^{m,n}_{S}P^{m,n}_{1}-P^{m,n}_{2}-P^{m,n}_{3})\right)\\
%&\;+\lambda_{R}\left(P_{R}-\sum^{N}_{m=1}\sum^{N}_{n=1}\eta^{m,n}_{R}P^{m,n}_{1}\right),
%%&+\underset{\{m,n\}\in S_{R}}\sum\rho_{m,n}\left(\gamma_{SR}^{m}P_{S}^{m,n}-\gamma_{SD}^{m}P_{S}^{m,n}-\gamma_{RD}^{n}P_{R}^{m,n}\right),
%\end{split}
%\end{equation}
where $\mu_{S}\geq0$, $\mu_{R}\geq0$ and
$\boldsymbol{\beta}=(\beta_{1},\beta_{2},...,\beta_{N})\succeq0$ are
dual variables. The dual objective function is
\begin{equation}
\begin{split}
g(\mu_{S},\mu_{R},\boldsymbol{\beta})=&\underset{\{\mathbf{P},\mathbf{t},\boldsymbol{\rho}\}}\max
L(\mathbf{P},\mathbf{t},\mu_{S},\mu_{R},\boldsymbol{\beta},\boldsymbol{\rho})\\
s.t.\quad &\mathbf{E}2,\; \mathbf{E}7, \;\mathbf{E}8,\;\mathbf{E}9,
%&P^{m,n}>0  \forall m,n,\\
%&t_{m,n}\geq0 \forall m,n,\\
%&\sum^{N}_{n=1}t_{m,n}=1 \forall m
\end{split}
\end{equation}
and the dual problem is
\begin{equation}  \label{Rd1}
\underset{\{\mu_{S},\mu_{R},\boldsymbol{\beta}\}}\min
g(\mu_{S},\mu_{R},\boldsymbol{\beta})\quad s.t.\;\;\mu_{S}\geq0,
\mu_{R}\geq0.
\end{equation}
We take derivatives of $L$ with respect to $P^{m,n}_{1}$, $P^{m,n}_{2}$ and $P^{m,n}_{3}$ and obtain
%Then we have
%\begin{equation}
%\frac{\partial L}{\partial
%S^{m,n}_{1}}=\frac{t_{m,n}\rho_{m,n}}{2}\frac{\overline{\gamma}^{m,n}_{1}}{t_{m,n}
%\rho_{m,n}+S^{m,n}_{1}\overline{\gamma}^{m,n}_{1}}-\alpha=0.
%\end{equation}
%Together with the constraint $P^{m,n}_{1}\geq0$, we obtain the
%optimal solution
\begin{equation}  \label{S1}
\begin{split}
P_{1*}^{m,n}&=t_{m,n}\rho_{m,n}\left[\frac{1}{2(\mu_{S}\eta^{m,n}_{S}+\mu_{R}\eta^{m,n}_{R})}
-\frac{1}{\overline{\lambda}^{m,n}}\right]^{+},\\
P_{2*}^{m,n}&=t_{m,n}(1-\rho_{m,n})\left[\frac{1}{2\mu_{S}}-\frac{1}{\lambda^{m}_{SD}}\right]^{+},\\
P_{3*}^{m,n}&=t_{m,n}(1-\rho_{m,n})\left[\frac{1}{2\mu_{R}}-\frac{1}{\lambda^{n}_{SD}}\right]^{+}.
\end{split}
\end{equation}
%where $[x]^{+}= \max\{0,x\}$. Similarly we obtain
%Then we take derivatives of $L$ with respect to $S^{m}_{2}$ and
%have
%\begin{equation}
%\frac{\partial L}{\partial
%S^{m}_{2}}=1+S^{m}_{2}\frac{\gamma^{m}_{SD}}{t_{m,n}(1-\rho_{m,n})}-\frac{\gamma^{m}_{SD}}{2\alpha}=0.
%\end{equation}
%Together with the constraint $S^{m}_{2}\geq0$, we obtain
%\begin{equation}  \label{S2}
%
%\end{equation}
%%Similarly we obtain the optimal solution of $S^{n}_{3}$ as
%\begin{equation}  \label{S3}
%
%\end{equation}
Denote $R_{m,n}^{R}$ and $R_{m,n}^{I}$ as the rate contribution of
$\mbox{SP}(m,n)$ to the Lagrange function in relaying mode and idle mode respectively. Then
\begin{equation}  \label{rr}
\begin{split}
&R_{m,n}^{R}=\frac{1}{2}\log
\left(1+\overline{\lambda}^{m,n}\tilde{P}_{1*}^{m,n}\right)
-\mu_{S}\eta^{m,n}_{S}\tilde{P}_{1*}^{m,n}\\
&\quad\quad\quad\;-\mu_{R}\eta^{m,n}_{R}\tilde{P}_{1*}^{m,n},\\
&R_{m,n}^{I}=-\mu_{S}(\tilde{P}_{2*}^{m,n}+\tilde{P}_{3*}^{m,n})\\
&\quad+\frac{1}{2}\biggl[\log
\left(1+\lambda^{m}_{SD}\tilde{P}_{2*}^{m,n}\right)+\log
\left(1+\lambda^{n}_{SD}\tilde{P}_{3*}^{m,n}\right)\biggr],
%\left[\frac{1}{2\alpha}-\frac{1}{\gamma^{m}_{SD}}\right]^{+}\right)
%-\alpha\left[\frac{1}{2\alpha}-\frac{1}{\gamma^{m}_{SD}}\right]^{+}\\
%&+\frac{1}{2}\log
%\left(1+\gamma^{n}_{SD}\left[\frac{1}{2\alpha}-\frac{1}{\gamma^{n}_{SD}}\right]^{+}\right)
%-\alpha\left[\frac{1}{2\alpha}-\frac{1}{\gamma^{n}_{SD}}\right]^{+}.
\end{split}
\end{equation}
%\begin{equation}  \label{ri}
%\begin{split}
%R_{m,n}^{NR}=&\frac{1}{2}\biggl[\log
%\left(1+\gamma^{m}_{SD}\tilde{P}_{2*}^{m,n}\right)+\log
%\left(1+\gamma^{n}_{SD}\tilde{P}_{3*}^{m,n}\right)\biggr]\\
%&-\lambda_{S}(\tilde{P}_{2*}^{m,n}+\tilde{P}_{3*}^{m,n}),
%%\left[\frac{1}{2\alpha}-\frac{1}{\gamma^{m}_{SD}}\right]^{+}\right)
%%-\alpha\left[\frac{1}{2\alpha}-\frac{1}{\gamma^{m}_{SD}}\right]^{+}\\
%%&+\frac{1}{2}\log
%%\left(1+\gamma^{n}_{SD}\left[\frac{1}{2\alpha}-\frac{1}{\gamma^{n}_{SD}}\right]^{+}\right)
%%-\alpha\left[\frac{1}{2\alpha}-\frac{1}{\gamma^{n}_{SD}}\right]^{+}.
%\end{split}
%\end{equation}
where $\tilde{P}_{1*}^{m,n}=\left[\frac{1}{2(\mu_{S}\eta^{m,n}_{S}+\mu_{R}\eta^{m,n}_{R})}
-\frac{1}{\overline{\lambda}^{m,n}}\right]^{+}$, $\tilde{P}_{2*}^{m,n}=\left[\frac{1}{2\mu_{S}}-\frac{1}{\lambda^{m}_{SD}}\right]^{+}$ and $\tilde{P}_{3*}^{m,n}=\left[\frac{1}{2\mu_{R}}-\frac{1}{\lambda^{n}_{SD}}\right]^{+}$.
Easily we obtain
\begin{equation}    \label{optr}
\rho^{*}_{m,n}=\left \{\begin{array} {c@{\;
\;}l} 1,\quad & \mbox{when} \quad R_{m,n}^{R}>R_{m,n}^{I},\\
0,\quad & \mbox{otherwise}.
\end{array}
\right.
\end{equation}
%
%\begin{equation} \label{optS}
%P_{1*}^{m,n}=t_{m,n}\tilde{S}^{m,n}=t_{m,n}\left[\frac{1}{2(\eta^{m,n}_{S}\lambda_{S}+\eta^{m,n}_{R}\lambda_{R})}
%-\frac{1}{\gamma^{m,n}}\right]^{+}.
%\end{equation}
Substitute~(\ref{S1}) into~(\ref{Lag1}) we obtain
\begin{equation}
\begin{split}
L(\mathbf{P},\mathbf{t},\mu_{S},\mu_{R},\boldsymbol{\beta},\boldsymbol{\rho})
= &t_{m,n}T_{m,n}+K_{\mu_{S},\mu_{R},\boldsymbol{\beta}},
\end{split}
\end{equation}
where $T_{m,n}=\rho_{m,n}R_{m,n}^{R}+(1-\rho_{m,n})R_{m,n}^{NR}-\beta_{n}$
%1/2\log(1+\gamma^{m,n}\tilde{S}^{m,n})-\lambda_{S}\eta^{m,n}_{S}\tilde{S}^{m,n}
%-\lambda_{R}\eta^{m,n}_{R}\tilde{S}^{m,n}-\beta_{n}$
and
$K_{\mu_{S},\mu_{R},\boldsymbol{\beta}}=\mu_{S}P_{S}+\mu_{R}P_{R}+\sum^{N}_{n=1}\beta_{n}$.
Both
$T_{m,n}$
and $K_{\mu_{S},\mu_{R},\boldsymbol{\beta}}$ are
independent of $t_{m,n}$. So the optimal $t^{*}_{m,n}$ is obtained as
\begin{equation}   \label{optt}
t^{*}_{m,n}=\left \{\begin{array} {c@{\; :
\;}l} 1 & m=\arg\underset{m=1,...,N}\max T_{m,n},\\
0 & otherwise.
\end{array} \forall n.
\right.
\end{equation}
$\mu_{S}$, $\mu_{R}$ and $\beta_{n}$ that
minimize $g(\mu_{S},\mu_{R},\boldsymbol{\beta})$ are achieved by the subgradient
method
\begin{equation}  \label{abc}
\left\{ {\begin{array}{lp{2mm}l}
   {\mu_{S}^{(i+1)}=\mu_{S}^{(i)}-a^{(i)}
   \Bigl(P_{S}-}\\
   {\quad\quad\quad\;\;\;\sum^{N}_{m=1}\sum^{N}_{n=1}(\eta^{m,n}_{S}P^{m,n}_{1(i)}-P^{m,n}_{2(i)}-P^{m,n}_{3(i)})\Bigr)}, \\
   {\mu_{R}^{(i+1)}=\mu_{R}^{(i)}-b^{(i)}
   \left(P_{R}-\sum^{N}_{m=1}\sum^{N}_{n=1}\eta^{m,n}_{R}P^{m,n}_{1(i)}\right)}, \\
   {\beta^{(i+1)}_{n}=\beta^{(i)}_{n}-c^{(i)}\left(1-\sum^{N}_{m=1}t^{(i)}_{m,n}\right),\; n=1,...,N.}  \\
\end{array} } \right.
\end{equation}
%\begin{equation}  \label{abc}
%\left\{ {\begin{array}{lp{2mm}l}
%   {\lambda_{S}^{(i+1)}=\lambda_{S}^{(i)}-a^{(i)}
%   \Bigl(P_{S}-}\\
%   \qquad\quad\;\;\;{\sum^{N}_{m=1}\sum^{N}_{n=1}(\eta^{m,n}_{S}P^{m,n}_{1(i)}-P^{m,n}_{2(i)}-P^{m,n}_{3(i)})\Bigr)}, \\
%   {\lambda_{R}^{(i+1)}=\lambda_{R}^{(i)}-b^{(i)}
%   \left(P_{R}-\sum^{N}_{m=1}\sum^{N}_{n=1}\eta^{m,n}_{R}P^{m,n}_{1(i)}\right)}, \\
%   {\beta^{(i+1)}_{n}=\beta^{(i)}_{n}-c^{(i)}\left(1-\sum^{N}_{m=1}t^{(i)}_{m,n}\right),\; n=1,...,N,}  \\
%\end{array} } \right.
%\end{equation}
By now, we have obtained the optimal mode selection vector, subcarrier pairing vector
$t^{*}_{m,n}$ as well as the power allocation vector
$\left(P_{1*}^{m,n},P_{2*}^{m,n},P_{3*}^{m,n}\right)$
for given dual variables
respectively. We can similarly update
the subcarrier pairing and power allocation vectors as in Algorithm~1 with some slight modifications.
The Lloyd algorithm can be employed again to design the codebook.

%=======================================================section4=====================================================================
\section{Simulation Results}\label{sec:4}

We present some simulations to demonstrate the
performance of the proposed algorithms in this section. The channels of the subcarriers are independent and
identically distributed (i.i.d.) subject to Rayleigh fading, with a large scale fading path loss exponent $2.5$. The channel coefficients are assumed
to be constant within two phases, and varying independently from one
period to another. We assume equal noise power at relay and
destination nodes, i.e., $\sigma_{r}^2=\sigma_{d}^2$.
In the simulations, QPSK modulation is adopted, and the
step sizes $a^{(i)}$ and $b^{(i)}$ for the subgradient method are
set to be $\frac{0.01}{\sqrt{i}}$, where $i$ is the iteration index. The size of CSI set in Algorithm~2 is $10^{4}$, which is far more than the quantized regions.
Several existing schemes are compared with the proposed algorithm in terms of sum rate. These existing schemes include:

\begin{enumerate}
\renewcommand{\labelenumi}{(\roman{enumi})}

\item \emph{UPA w/o SP}: the messages transmitted on subcarrier $m$ at
the source node will be retransmitted on the subcarrier $m$ at the
relay node; the power is allocated equally at the source and relay
subcarriers.
\item \emph{OPA w/o SP}: the messages transmitted on the subcarrier $m$
at the source node will be
retransmitted on the subcarrier $m$ at the relay node; the power
allocation is performed according to water-filling at the source and the
relay subcarriers.
\item \emph{UPA with SP}: the messages transmitted on the subcarrier $m$ at the
source node will be retransmitted on the subcarrier $n$ at the relay
node by subcarrier pairing; the power is allocated equally at the
source and relay subcarriers.
\end{enumerate}
Then the performance of the proposed algorithm with different feedback bits are demonstrated. Besides, the performance gap between the
\emph{enhanced} DF and the \emph{selective} DF modes versus the
subcarrier number is also revealed in our simulations.

%\begin{figure}[!t]
%\centering
%\includegraphics[width=3.3in,angle=0]{fig1.eps}

%\end{figure}
\subsection{Rate Comparison for Different Schemes}

Schemes
(\setcounter{fofo}{1}\roman{fofo})-(\setcounter{fofo}{3}\roman{fofo})
are compared with the proposed algorithm with perfect CSI and
limited feedback scheme in Fig.~2. The upper curve denotes the proposed
joint power allocation and subcarrier pairing for the \emph{enhanced} DF
scheme with perfect CSI. The second upper curve denotes the proposed
joint power allocation and subcarrier pairing for the \emph{enhanced} DF
scheme with 2 bits feedback. The other three curves denote the
existing schemes
(\setcounter{fofo}{1}\roman{fofo})-(\setcounter{fofo}{3}\roman{fofo})
respectively. We can observe that, only with 2 bits feedback, the proposed
joint power allocation and subcarrier pairing for the \emph{enhanced} DF
relay outperforms the existing schemes
(\setcounter{fofo}{1}\roman{fofo})-(\setcounter{fofo}{3}\roman{fofo}) greatly.
So we can conclude that the joint power allocation and subcarrier
pairing make valuable contribution to system sum rate.
\begin{figure}[!t]
\centering
\includegraphics[width=3.5in,angle=0]{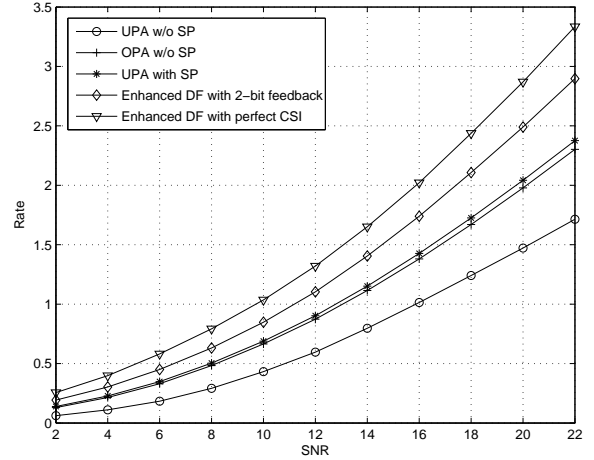}
\caption{System sum rate versus SNR for the proposed enhanced DF
relay schemes and the existing schemes. Where "w/o" denotes
"without".} \label{fig2}
%the upper curve denotes dual problem solution, the lowest curve
%shows system rate without SP and optimal power allocation. The other
%four curves denote schemes $(ii)$, $(iii)$ and the proposed modified
%idle scheme with 2-bit feedback.
\end{figure}

\subsection{Rate Comparison for Different Feedback Bits}

The joint power allocation and subcarrier pairing for the
\emph{enhanced} DF relay with different feedback bits are compared
in Fig.~3. We can find that only a few feedback bits are enough to
achieve most of the performance gain of the perfect feedback. For
example, with $4$ bits of feedback at rate of $2.5$ in Fig.~3, there
is only a $-1.7dB$ gap to the perfect CSI case, and we also notice
that further increasing the feedback bits bring degressive
improvement, which implies that the feedback bits as well as the
codebook size in the model are not necessarily too large.
\begin{figure}[!t]
\centering
\includegraphics[width=3.5in,angle=0]{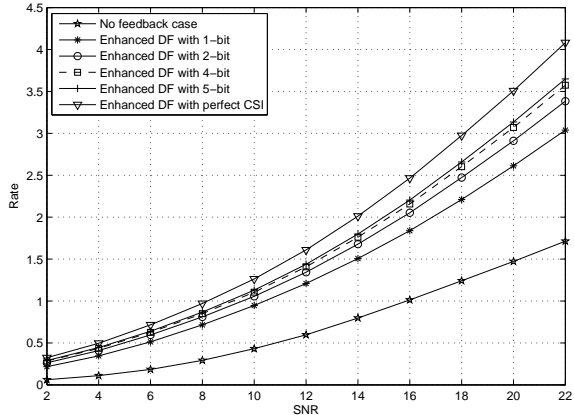}
\caption{System sum rate versus SNR for the proposed enhanced DF
relay scheme with different levels of feedback bits. The upper curve
denotes the perfect CSI case, the lowest curve denotes the scheme
without feedback, where the power is uniformly allocated. Others
curves demonstrate the effect of different feedback bits on sum
rate.} \label{fig2}
\end{figure}

\subsection{Rate Versus Different Subcarrier Numbers Under Sum and Individual Power Constraints}

With the sum and individual power constraints, the sum rates versus
the number of subcarriers for the \emph{selective} DF and the
\emph{enhanced} DF relaying modes are illustrated in Fig.~4 and
Fig.~5. We consider the cases that subcarrier number $N=2,4,8,16$
with fixed feedback bit level of $2$. For the sake of fairness, we
assume $P_{S}+P_{R}=P_{t}$. In addition, as for the case with
individual power constraints, we assume $P_{S}=3P_{R}$. The
constraints are set with the practical consideration that the relay
often plays the role of assisting the transmission between the
source and the destination. Moreover, if more power is assigned to
the relay node, the achievable rate will be limited since some of
the relay power will not be used. Assuming $P_{S}=\frac{3}{4}P_{t}$
will make the comparison with the sum power constraints much fairer.
Besides, we assume that the relay locates in a line between the
source and the destination and the SD distance is one unit. Denote
$d$ $(0 < d < 1)$ as the SR distance. Thus the RD distance is $1-d$.
Fig.~4 is obtained with $d=0.4$, while $d=0.8$ in Fig.~5. We find
that the \emph{enhanced} DF mode always outperforms the
\emph{selective} DF mode and the schemes without subcarrier pairing,
especially when the channel condition of RD is relatively poor. We
can also observe that the bigger the subcarrier number is, the
bigger the performance gap between the two modes is. As for the
cases under different constraints, the performance of the sum power
constraint is better than that of the individual power constraints,
which is due to the more flexibility of power allocation between source
and relay under the sum power constraint. Besides, we consider the
duality gaps in the two figures. The simulation results exactly
coincide with our analysis in
section~\setcounter{fofo}{3}\Roman{fofo}. We find that the dual
solutions approximate to the optimal values of~(\ref{ori}) in our
simulation. The duality gaps turn out to be nearly $0$ when the
number of subcarriers is reasonably large, which is consistent with
the prediction in~\cite{IEEEconf:44}.
\begin{figure}[!t]
\centering
\includegraphics[width=3.5in,angle=0]{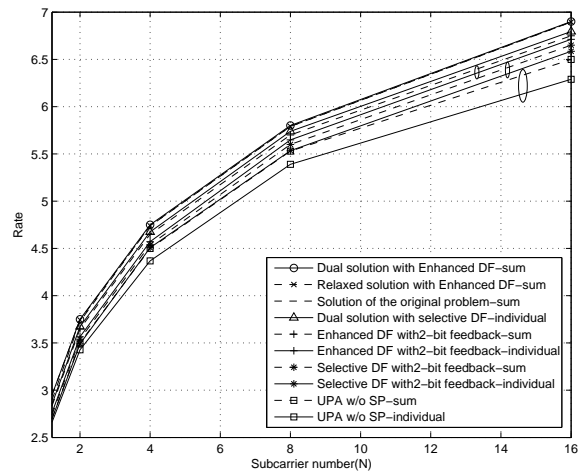}
\caption{The sum rate versus the number of subcarriers for the
enhanced DF, selective DF with fixed feedback bit of $2$. These
curves are obtained with $d=0.4$.} \label{fig3}
\end{figure}
\begin{figure}[!t]
\centering
\includegraphics[width=3.5in,angle=0]{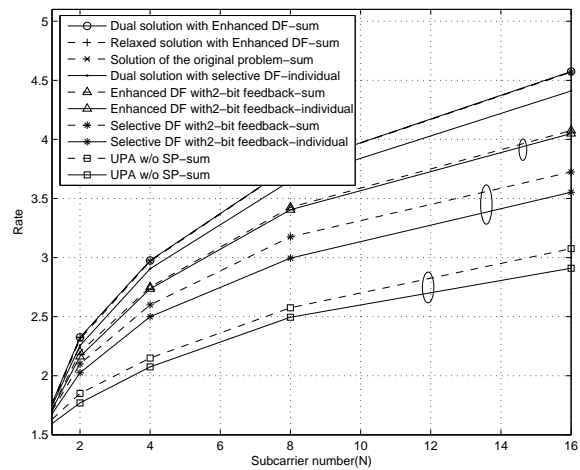}
\caption{The sum rate versus the number of subcarriers and the
performance gap between the schemes with modified idle and selective
relaying modes. We assume that system operates with fixed feedback bit
level of 2. The curves are obtained with $d=0.8$.} \label{fig4}
\end{figure}

\subsection{The Effect of Relay Location On Rate}

In order to exploit the system rate versus SNR for different relay
locations, we simulate the rate versus SNR by setting $d=0.25$,
$d=0.5$ and $d=0.75$ respectively. Fig.~6 demonstrates the effect of
relay location to system sum rate at different SNR with a fixed
feedback bit level of $2$. We can find that the \emph{enhanced} DF
mode always outperforms the \emph{selective} DF mode and the
\emph{OPA w/o SP} in any kind of $d$, and the channel condition of
SR plays a more important role than the channel condition of RD in
general.
%is more the better the
%channel condition $S$-$R$ is, the more excellent system performance
%is.
Besides, We find that the gap between the system sum rates achieved
by the \emph{enhanced} DF and the \emph{selective} DF is larger when $|d-0.5|$ is
larger; while the performance gap between the \emph{enhanced} DF and the
\emph{selective} DF is tiny when $d=0.5$.

\begin{figure}[!t]
\centering
\includegraphics[width=3.5in,angle=0]{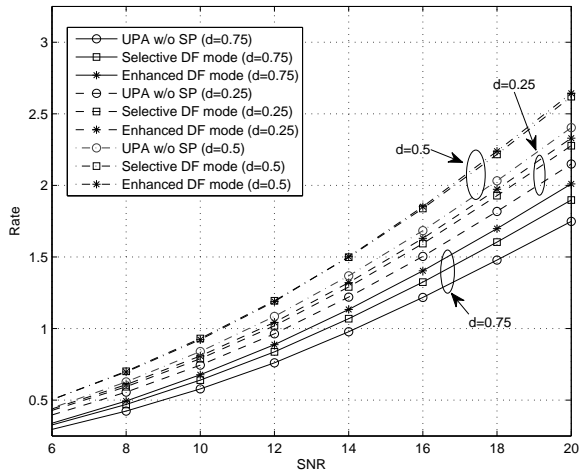}
\caption{The sum rate versus average SNR for
$d=0.25,\,0.5,\,0.75$, with feedback bits of $2$.} \label{fig5}
\end{figure}

\begin{figure}[!t]
\centering
\includegraphics[width=3.5in,angle=0]{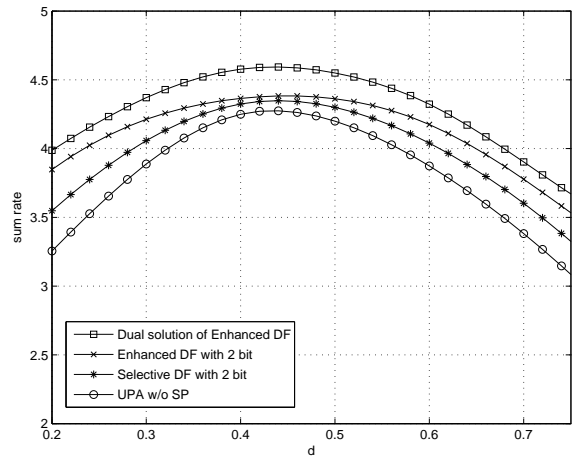}
\caption{The sum rate versus the relay location for the enhanced DF
and selective DF with $2$-bit feedback when N=4.} \label{fig5}
\end{figure}

To exploit the effect of relay location to the system
performance, we
simulate the sum rate versus relay location $d$ in Fig.~7. The figure is obtained with the
fixed feedback bit $2$ and subcarrier number $N=4$. We find that the
rate reaches maximum at about $d=0.45$. We also observe that
comparing with the proposed scheme with $2$-bit feedback, the
performance loss of the scheme $\emph{UPA with SP}$ is not very big
at this location, which implies that if none of SR or RD channels is
very poor, or there is no great difference between the channel
conditions of SR and RD, and the scheme $\emph{UPA with SP}$ can
provide acceptable performance with $N=4$. But if at least one of
these channel conditions is very poor, we'd better dynamically
allocate the power and subcarrier resources, since the proposed
algorithm can achieve remarkable performance gain. In addition, the
performance loss of scheme $\emph{UPA with SP}$ increases with the
number of subcarriers due to frequency diversity and more
flexibility in pairing of large $N$. Fig.~8 and Fig.~9 are obtained
with $N=32,64$ respectively. We observe that the performance gains of
the proposed algorithm are much more remarkable. The remarkable
performance gain results from much more pairing degree provided by
the big subcarrier number. There is another general trend can be
observed from the two figures. The rate gap between the
\emph{enhanced} DF and the \emph{selective} DF is larger when
$|d-0.5|$ is larger. The performance improvement of the
\emph{enhanced} DF is due to the extra direct-link transmission in
the second phase, since the relay has high possibility to be idle
when $S$-$R$ or $R$-$D$ channel is poor because of the large
$|d-0.5|$.
\begin{figure}[!t]
\centering
\includegraphics[width=3.5in,angle=0]{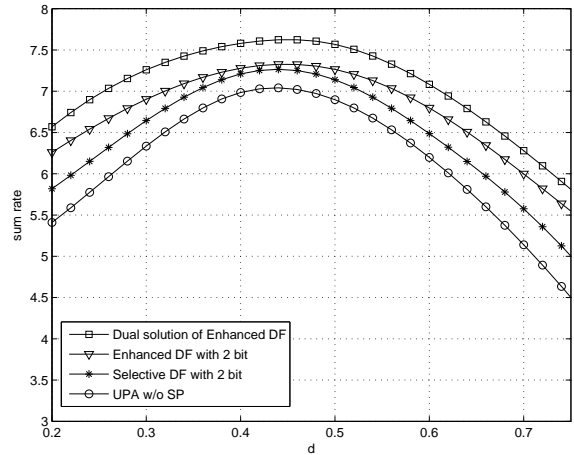}
\caption{The sum rate versus the relay location for the enhanced DF
and selective DF with $2$-bit feedback when N=32.} \label{fig5}
\end{figure}

\begin{figure}[!t]
\centering
\includegraphics[width=3.5in,angle=0]{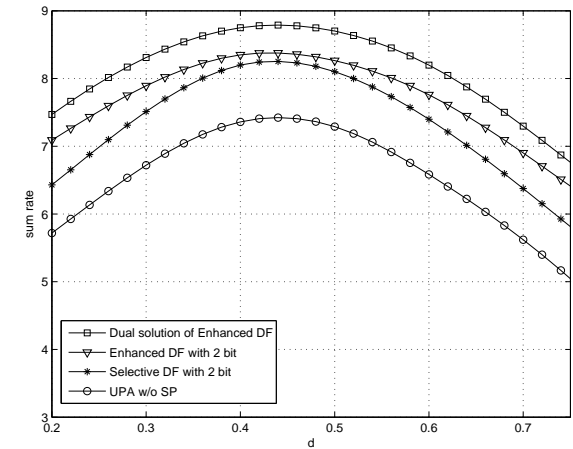}
\caption{The sum rate versus the relay location for the enhanced DF
and selective DF with $2$-bit feedback when N=64.} \label{fig5}
\end{figure}

%=======================================================section5=====================================================================
\section{Conclusion}\label{sec:5}

In this paper, we discuss a limited feedback based joint power
allocation and subcarrier pairing algorithm for the OFDM DF relay networks with diversity.
When the relay does not
forward the received symbols on some subcarriers, we further allow the source node to transmit
new messages on these idle subcarriers.
Both sum power constraint and individual
power constraints for the source and relay nodes are considered. Since the formulated optimization is a mixed integer programming problem, we transform it into a convex problem by continuous relaxation and then solve it in the dual domain. Simulations show that the proposed
algorithms can achieve considerable rate gain with tractable complexities. It outperforms several existing schemes under various channel
conditions. The contribution of the extra
direct-link transmission is also clearly demonstrated in the simulation.
In addition, we notice that a negligible performance loss can be
achieved with just a few feedback bits at different levels of SNR
values.
\end{document}